\documentclass[prd,byrevtex
,preprint
,tightenlines
,showkeys
,showpacs
,nofootinbib
,amsfonts,amsmath,amssymb
]{revtex4}
\usepackage[pdftex]{graphicx}
\usepackage{float}
\usepackage{color}
\newcommand{\al}{\alpha}
\newcommand{\pa}{\partial}
\newcommand{\del}{\delta}
\newcommand{\anno}[1]{#1}

\begin{document}
\title{Wave optics in spacetimes with compact gravitating object}
\author{Yasusada Nambu}
\email{nambu@gravity.phys.nagoya-u.ac.jp}
\affiliation{Department of Physics, Graduate School of Science, Nagoya
University, Chikusa, Nagoya 464-8602, Japan}

\author{Sousuke Noda}
\email{sousuke.noda@yukawa.kyoto-u.ac.jp}
\affiliation{\anno{Center for Gravitation and Cosmology, College of
    Physical Science and Technology, Yangzhou University, Yangzhou
    225009, China} \\and \\ Yukawa Institute for Theoretical Physics, 
Kyoto University, Kitashirakawa Oiwakecho Sakyoku, Kyoto 606-8502, Japan}

\author{Yuichiro Sakai}
\affiliation{Department of Physics, Graduate School of Science, Nagoya
University, Chikusa-ku, Nagoya 464-8602, Japan}

\date{September 17, 2019} 

%
\begin{abstract} 
  We investigate the wave optics in spherically symmetric spacetimes: 
  Schwarzschild black hole, spherical star with a perfect absorbing 
  surface, and massless/massive Ellis wormholes. 
  Assuming a point wave source, wave patterns and power
  spectrums for scattering waves are obtained by solving the scalar
  wave equation numerically.  We found that the power spectrum
    at the observer in the forward direction shows oscillations with
    two characteristic periods determined by the interference effect
    associated with the photon sphere and the diffraction effect due
    to the absorbing boundary condition inside of the photon sphere.
\end{abstract}   

\maketitle

\tableofcontents

\newpage
\section{Introduction}

The photon sphere is a set of circular unstable photon orbits around a
gravitating object and it \anno{forms} a two-dimensional sphere with a
constant radius for spherically symmetric static spacetimes.
Recently, related to the existence of the photon sphere, bright ring
and the ``shadow'' of M87 have been observed~\cite{EHT}.
The properties of shadows of strong gravitating objects such as
black holes have been studied in detail~                                        
\cite{Cunha2017,Cunha2017a,Cunha2017b,Cunha2018,Cunha2018a,Cunha2018b,Sakai2014,Ohgami2015,Ohgami2016}.
\anno{The shadow is a dark region on the observer's sky and its rim
  corresponds to the photon sphere projected onto the observer's sky.}
 By its definition, information inside of the photon
sphere cannot be detectable by light rays unless an illuminating light
source is placed inside of the photon sphere.

Although the photon sphere is introduced in terms of null geodesics,
which are rays in the geometrical optics, the relation to the
quasinormal modes of the black hole has been also discussed so
far~\cite{Cardoso2009}.  The quasinormal modes of black holes are
obtained as poles of the scattering matrix in the complex frequency
domain and its eikonal limit corresponds to light rays of the unstable
photon orbits around black holes. Based on the established treatment of wave
scattering problems (partial wave decomposition, phase shift etc.; see
Refs.~\cite{Andersson2000,Glampedakis2001}), the
photon sphere is related to Regge poles which are poles of the
scattering matrix in the complex angular momentum space. Thus it is
possible to understand properties of spacetimes with strong
gravitating objects using wave optics.  As an application to this
direction, imaging of black hole photon sphere with waves was
investigated by Kanai and Nambu~\cite{Kanai2013} and Nambu and
Noda~\cite{Nambu2016}.  The reconstruction of black hole images from
scattering waves was attempted by Fourier transform of scattered
waves.  Other approaches to the wave scattering by black holes
  such as the evaluation of the differential cross section have been
  investigated by many authors
  \cite{Matzner1968,Sanchez1978,Handler1980,Zhang1984,Matzner1985,Futterman,Dolan2008a,Dolan2008,Crispino2009,Crispino2014,Crispino2015,Leite2017,Alexandre2018,Sporea2018}.
  Recently, wave scattering by stars is also discussed
  \cite{Dolan2017,Stratton2019,Cotaescu2019}.

Concerning the wave optical effect for the weak gravitational lensing,
interference fringe patterns in the spatial domain (scattering
amplitude) and the frequency domain (power spectrum) are
expected. They are caused by interference between two coherent light
rays (direct rays). For the gravitational lensing by a black hole, an
additional interference effect associated with the photon sphere is
expected. Light rays can go around the black hole an arbitrary number
of times (orbiting), the direct rays and these winding rays can
interfere, and an additional component of fringe appears in the power
spectrum.  In the paper~\cite{Nambu2016}, the analytic expression for
scattering waves by the Schwarzschild black hole was derived in the
eikonal limit (the leading order of the wave effect), and wave optical
images of Einstein rings and the photon sphere were
obtained. Moreover, modulation of power spectrums caused by the photon
sphere was also clarified. As an astrophysical application of wave
optics to gravitational lensing systems, Yoo \textit{et
  al.}~\cite{Yoo2013} investigated the behavior of power spectrums from a
point source and discussed the possibility to distinguish Ellis wormhole
spacetimes from spacetimes with a point mass. Their analysis is based
on the weak field approximation of the gravitational field (weak lensing
effect). They concluded that the Ellis wormhole spacetime shows
different behaviors of the power spectrum due to the $r^{-2}$ law of the
wormhole's gravitational potential. However, their analysis lacks
strong lensing effect associated with the photon sphere.

In this paper, we consider wave optical properties of spacetimes with
the photon sphere. Let us consider a situation where a wave source is
located outside of the photon sphere of the gravitating object and an
observer detects a scattered signal. In the geometrical optics, as light
rays captured by the photon sphere cannot escape from it, we cannot
look inside of the photon sphere using a light source placed outside
of the photon sphere\anno{,} if objects inside of the photon sphere do
not emit and reflect light rays. However, in the wave optics, even if
a part of the wave propagates inside of the photon sphere, it can escape
to outside due to the wave effect and \anno{it is possible to }extract
information \anno{of interior} of the photon sphere.  This expectation is
directly connected to the discrimination problem of gravitating
objects called black hole mimickers such as ultracompact objects
\anno{using wave optical effects. The black hole mimickers have photon
  spheres but no event horizons. Thus it is not possible to
  discriminate between the black hole mimickers and black holes using
  light rays.  In this paper, as models of the black
hole mimickers}, we consider a spherical star with a perfect absorbing
surface and massless/massive Ellis wormholes.
We mainly focus on behavior of power spectrums in the forward scattering case;
the path difference between two direct rays is zero and if the
gravitating object has no structure like the photon sphere, 
we do not have any interference fringe in the power spectrum. 
However, if the gravitating object has the photon sphere or some structures, 
 modulations \anno{of} the power spectrum caused by interference between direct 
rays and orbiting rays \anno{are} expected.  

The structure of the paper is as follows. In Sec.~II, we introduce our
setup of the wave scattering problem in spherically symmetric spacetimes
and explain our numerical methods. We present our numerical results in
Sec.~III for the Schwarzschild spacetime and the Ellis wormhole
spacetime. In Sec.~IV, we apply a formula in the wave optics to
explain \anno{the} interference \anno{fringe} that appeared in power
spectrums. Section V is devoted to the summary and conclusion.

\newpage
\section{Wave optics in static spherically symmetric  spacetimes}

In this section, we introduce the setup of the problem and numerical
method to obtain \anno{scattering waves} by a spherical gravitating object.
\subsection{Wave equation with a point source}
\begin{figure}[H]
  \centering
  \includegraphics[width=0.6\linewidth,clip]{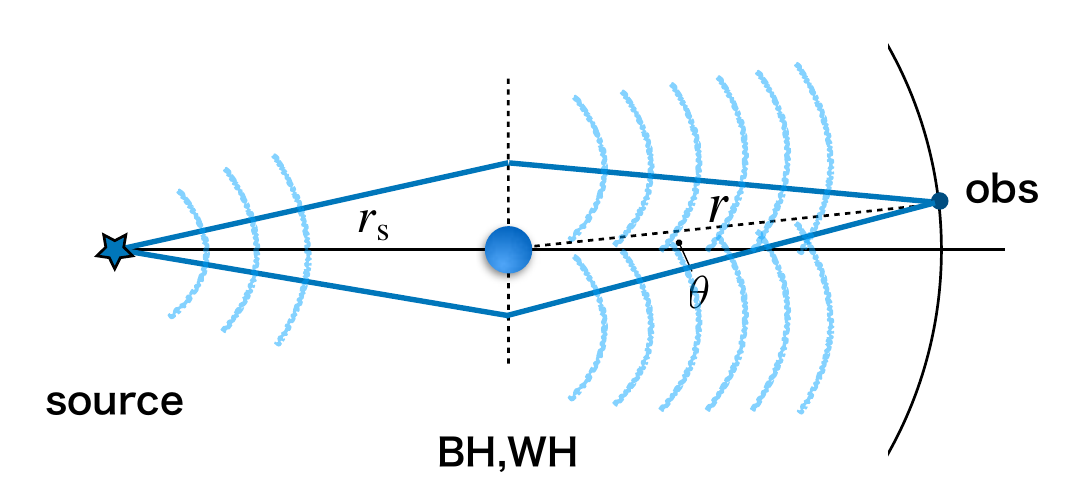}
  \caption{Configuration of our wave scattering problem. An
      observer receives waves from a point wave source. } 
  \label{fig:setup}
\end{figure}
Figure \ref{fig:setup} shows our setup of the wave scattering problem.
We consider a massless scalar field as the benchmark treatment for
wave scattering problems and we do not consider polarization degrees
of freedom that are necessary for the electromagnetic and
gravitational waves. The background geometry is assumed to be 
static and spherically symmetric spacetimes with the metric
\begin{equation}
  ds^2\anno{=g_{\mu\nu}dx^\mu dx^\nu}=-f(r)dt^2+\frac{dr^2}{\anno{h(r)}}+r^2 d\Omega^2.
\end{equation}
For a monochromatic stationary wave with time dependence
$e^{-i\omega t}$, the wave equation for the massless scalar field
\anno{$\Phi$} reduces to the following Helmholtz type equation with a
source term
\begin{equation}
-g^{00}\omega^2\Phi+\frac{1}{\sqrt{-g}}\pa_j\left(\sqrt{-g}\,g^{jk}\pa_k\Phi
\right)
  =-\anno{S_\omega}\,\del^3(\vec{r},\vec{r}_s),\quad i,j,k=r,\theta,\phi,
\end{equation}
\anno{where a point wave source is placed at $\vec{r}_s$ and
  $\del^3(\vec{r},\vec{r}_s)$ is the invariant delta function
  $\frac{1}{\sqrt{-g}}\del^3(\vec{r}-\vec{r}_s)$.} \anno{$S_\omega$
  denotes the Fourier amplitude of the wave source. In this paper, we
  assume the spectrum of the wave source has no $\omega$ dependence
  and $S_\omega$ is $\omega$ independent constant. The power spectrum
  of the wave at observing point is given by
  $|\Phi(\omega)|_\text{obs}^2$.}\footnote{\anno{Strictly speaking,
    $|\Phi(\omega)|^2$ represents energy of the wave per unit interval
    of the frequency and the power of the wave for interval of the
    frequency $\Delta\omega$ is represented as
    $|\Phi(\omega)|^2\Delta\omega$. In this paper, we call
    $|\Phi(\omega)|^2$ as the power spectrum for simplicity.}}
\anno{We assume the wave source is placed on the $-z$ axis: $r=r_s, \theta=\pi$.
  That is
  $\delta^3(\vec{r},\vec{r_s})\propto\delta(r-r_s)\delta(\cos\theta+1).$}
Owing to the symmetry of the spacetime, the wave function can be
separated as
\begin{equation}
  \Phi(r,\theta)=\frac{1}{r}\sum_{\ell=0}^{\infty}R_{\ell}(r)
  P_\ell(\cos\theta), 
  \label{eq:el-sum}
\end{equation}
and \anno{using the formula
  $\delta(\cos\theta+1)=\sum_{\ell=0}^\infty(-)^\ell(\ell+1/2)P_\ell(\cos\theta)$},
the radial wave function $R_\ell$ obeys the following Schr\"{o}dinger
type equation
\begin{equation}
  \frac{d^2R_\ell}{dx_\text{tot}^2}+\left(\omega^2-V_\text{eff}\right)R_\ell=a_s(-)^\ell(\ell+1/2)\,\del(r-r_s),
  \label{eq:radial-eq}
\end{equation}
where  $a_s$ denotes the \anno{$\omega$ independent} amplitude of the
point source, the tortoise coordinate is introduced \anno{as}
\begin{equation}
  x_\text{tot}=\int\frac{dr}{\sqrt{f\anno{h}}},
\end{equation}
and the effective potential is defined by
\begin{equation}
  V_\text{eff}=f\frac{\ell(\ell+1)}{r^2}+\frac{(f\anno{h})_{,r}}{2r}.
\end{equation}

In this paper, we consider the Schwarzschild spacetime and the Ellis
wormhole spacetime. The metric of the Schwarzschild spacetime with the
tortoise coordinate $x_\text{tot}$ is
\begin{align}
  &ds^2_\text{Schw}=\left(1-\frac{2M}{r}\right)(-dt^2+dx_\text{tot}^2)+r^2
    d\Omega^2,\\
  & x_\text{tot}=r+2M\ln\left(\frac{r}{2M}-1\right),\quad -\infty<x_\text{tot}<+\infty.
\end{align}
The effective potential is
\begin{equation}
  V_\text{eff}=\left(1-\frac{2M}{r}\right)\left(\frac{\ell(\ell+1)}{r^2}+\frac{2M}{r^3}\right). 
  \label{eq:pot-BH}
\end{equation}
The metric of the Ellis wormhole (massless case) is
\begin{align}
  &ds^2_\text{WH}=-dt^2+dx^2+r^2 d\Omega^2,\\
  &r=\sqrt{x^2+a^2},\quad -\infty<x<+\infty,
\end{align}
where the parameter $a$ represents the size of the wormhole's throat.  The
effective potential is
\begin{equation}
  V_\text{eff}=\frac{\ell(\ell+1)}{r^2}+\frac{a^2}{r^4}.
\end{equation}
\begin{figure}[H]
  \centering
  \includegraphics[width=0.5\linewidth,clip]{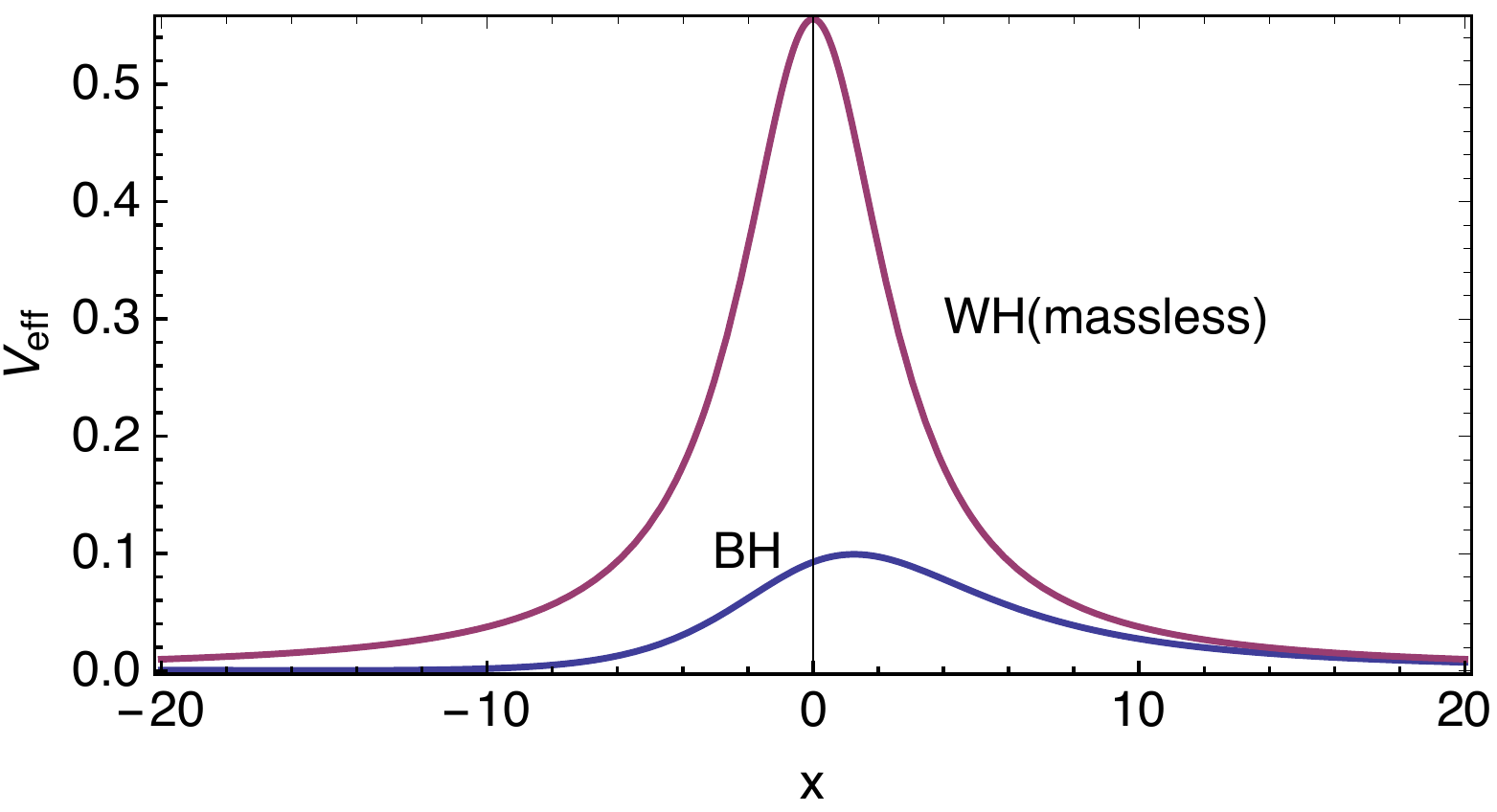} 
  \caption{Effective potentials for the Schwarzschild
    spacetime and the Ellis wormhole (massless) spacetime. The plot is with
    $\ell=2, a=3M$. The circumference radius of the
    photon sphere for both spacetimes is $3M$.  }
\end{figure}
\noindent
The metric for the massive Ellis wormhole is presented in Appendix A
[Eq.~\eqref{eq:metric-wh}]. Figure 2 shows these effective potentials.
\subsection{Our numerical methods}
We present here the numerical method adopted in our analysis.  We first
obtain the solution of the radial equation \eqref{eq:radial-eq}
numerically. We impose two boundary conditions at $r=r_\text{in}$ (an
inner boundary corresponds to the black hole horizon, the star's surface,
and another asymptotic flat region of wormhole) and $r=r_\text{out}$ (an
outer boundary corresponds to the spatially far region). We prepare
two solutions of the homogeneous radial equation without a source
term:
    \begin{align*}
      &u_1(r),\quad r\in[r_\text{in}, r_s),\quad \text{BC is imposed at
        $r_\text{in}$}, \\
      &u_2(r),\quad r\in(r_s,r_\text{out}],\quad \text{BC is imposed
        at $r_\text{out}$}
    \end{align*}
where $u_1$ is obtained by integrating the radial equation from $r_\text{in}$
to $r_s$, and $u_2$ is obtained by integrating the radial equation from
$r_\text{out}$ to $r_s$.

 Radial functions $u_1$ and $u_2$ do not satisfy the boundary
condition at the source $r_s$ which is obtained by integrating
\eqref{eq:radial-eq} around $r_s$:
    \begin{equation}
      \left(\frac{dR_2}{dr}-\frac{dR_1}{dr}\right)_{r=r_s}=a_s(-)^\ell
      (\ell+1/2)\equiv \Delta_\ell,\quad (R_2-R_1)_{r=r_s}=0,
      \label{eq:matching}
    \end{equation}
where $R_1(r)=R(r\leq r_s)$ and $R_2(r)=R(r\geq r_s)$.
    Using $u_1$ and $u_2$, we introduce new radial functions as
    \begin{equation}
      R_1=c_1u_1,\quad R_2=c_2u_2,
    \end{equation}
    where $c_1$ and $c_2$ are constants to be determined by the matching
    condition \eqref{eq:matching} at $r_s$:
    \begin{equation}
      c_2u_2'(r_s)-c_1u_1'(r_s)=\Delta_\ell,\quad c_1u_1(r_s)=c_2u_2(r_s).
    \end{equation}
We obtain
\begin{equation}
  c_1=\left(\frac{u_2}{W[u_1,u_2]}\right)_{r=r_s}\Delta_\ell,\quad
  c_2=\left(\frac{u_1}{W[u_1,u_2]}\right)_{r=r_s}\Delta_\ell.
\end{equation}
where $W=u_1u_2'-u_2u_1'$ is the Wronskian. Thus, $R_1$ and $R_2$
\anno{with required boundary conditions} are
\begin{equation}
  R_1=\frac{\Delta_\ell}{W[u_1,u_2]_{r_s}}u_1(r)u_2(r_s),\quad
  R_2=\frac{\Delta_\ell}{W[u_1,u_2]_{r_s}}u_1(r_s)u_2(r).
\end{equation}

To obtain numerical solutions $u_1$ and $u_2$, we adopted the fourth-order
Runge-Kutta method. We obtained the radial wave function within the
relative errors $10^{-5}$.  Concerning the value $\ell_\text{max}$ for
the summation of the partial waves \eqref{eq:el-sum}, we determined it by
checking the convergence of $\Phi$ at $r_\text{obs}=20M, \theta=0$. For
$M\omega<1$, \anno{we obtained} $\ell_\text{max} =10-13$. For
$1<M\omega\leq 10$, we found that $\ell_\text{max}=9\times(M\omega+1)$
for the black hole cases. For calculations with stars and a wormhole, we
determined $\ell_\text{max}$ using the same method.

\section{Results}
We obtained the scattering wave from a monochromatic point wave source
for the Schwarzschild spacetime (black hole, stars with a perfect
absorbing surface) and the Ellis wormhole spacetime.  \anno{The} point
\anno{wave} source is placed at $r_s=6M,~\theta=\pi$ with frequencies
$0< M\omega\leq 10$. The observing point of power spectrum is
$r_\text{obs}=20M$.

\subsection{Black hole case} 
Numerical results for the Schwarzschild spacetime \anno{are} as follows.
\begin{figure}[H]   
  \centering
    \includegraphics[width=1.\linewidth,clip]{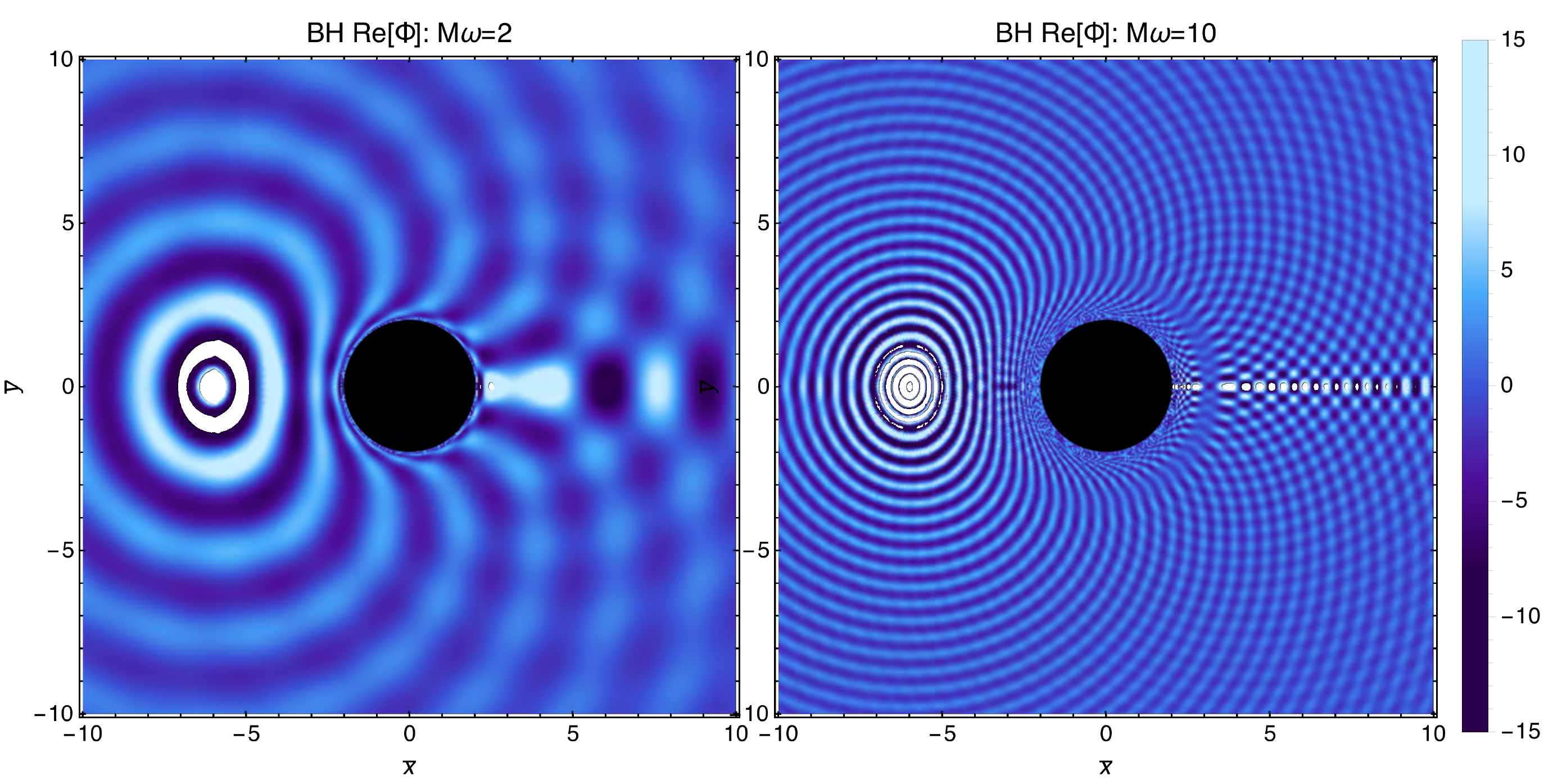}
  \caption{Real part of $\Phi$ for  $M\omega=2,10$. The Cartesian 
    coordinates $\bar x$ and $\bar y$ are introduced by $\bar x=r\cos\theta,
    \bar y=r\sin\theta$.}
  \label{fig:BH-wavepattern}
\end{figure}
\noindent
Figure~\ref{fig:BH-wavepattern} shows the real part of scattering waves.
For $M\omega =10$, we can see the circlelike wave pattern
corresponding to the photon sphere at $r\simeq 3M$ and the bright line
(caustics) behind the black hole,  while, for the $M\omega =2$ case, these
features are blurred due to the wave effect.
Figures~\ref{fig:power-BH}, \ref{fig:power3-BH} and \ref{fig:power2-BH}
show the intensity of scattered waves at $r_\text{obs}$.  They show
interference fringes in both the spatial domain ($\theta$) and the frequency
domain ($\omega$).  Namely, \anno{ $|\Phi|^2$ on a constant-$\theta$
  slice} is the power spectrum at the \anno{observing} point and
\anno{$|\Phi|^2$ on a constant-$M\omega$ slice} represents the
scattering amplitude for \anno{a} fixed frequency.
\begin{figure}[H] 
  \centering 
    \includegraphics[width=0.9\linewidth,clip]{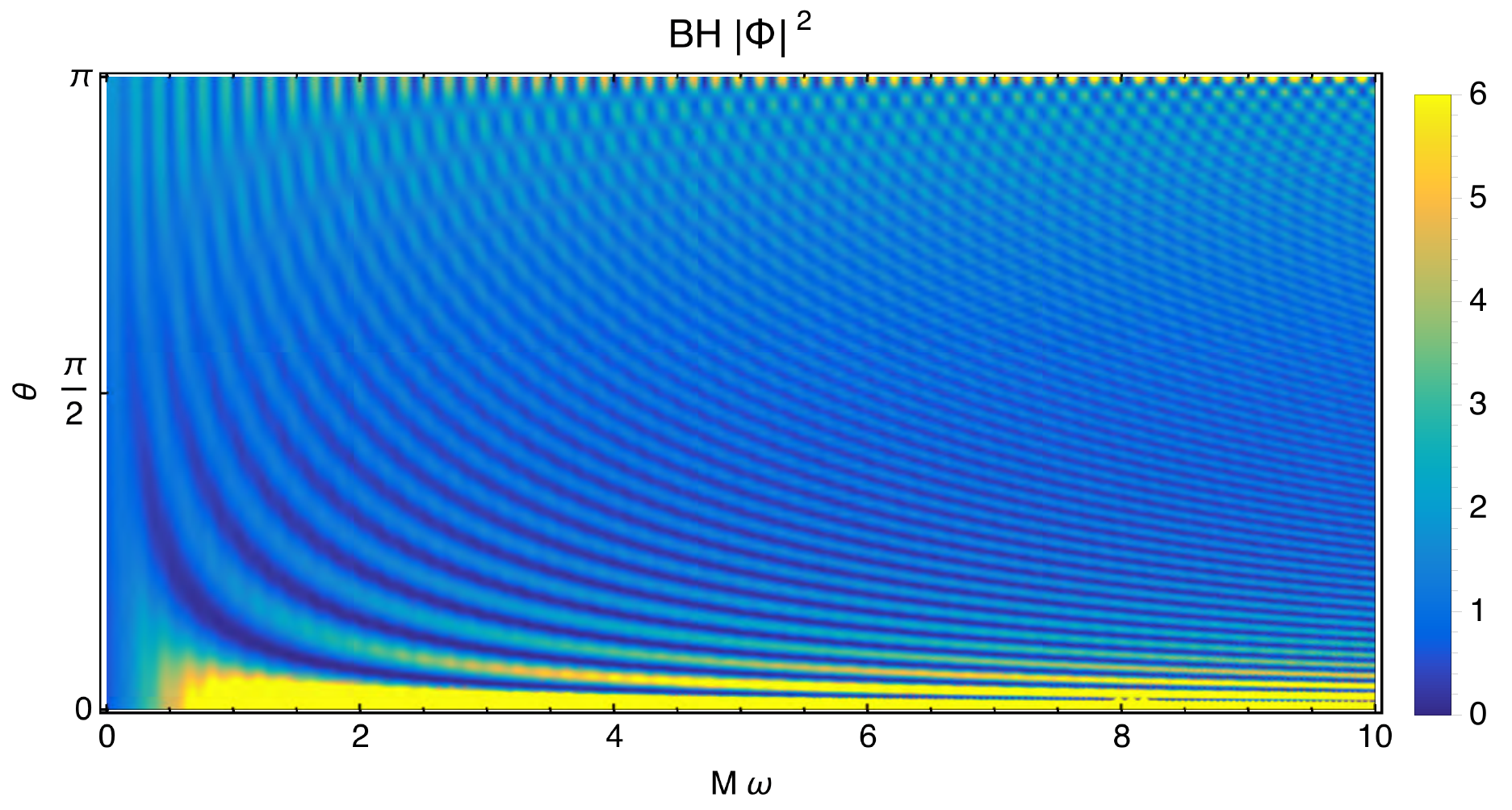}
    \caption{$|\Phi|^2$ at $r_\text{obs}$ as a function of
      $(\omega, \theta)$. Interference fringe appears in the
      two-dimensional parameter space.  }
  \label{fig:power-BH} 
\end{figure}
 \begin{figure}[H] 
   \centering
    \includegraphics[width=1\linewidth,clip]{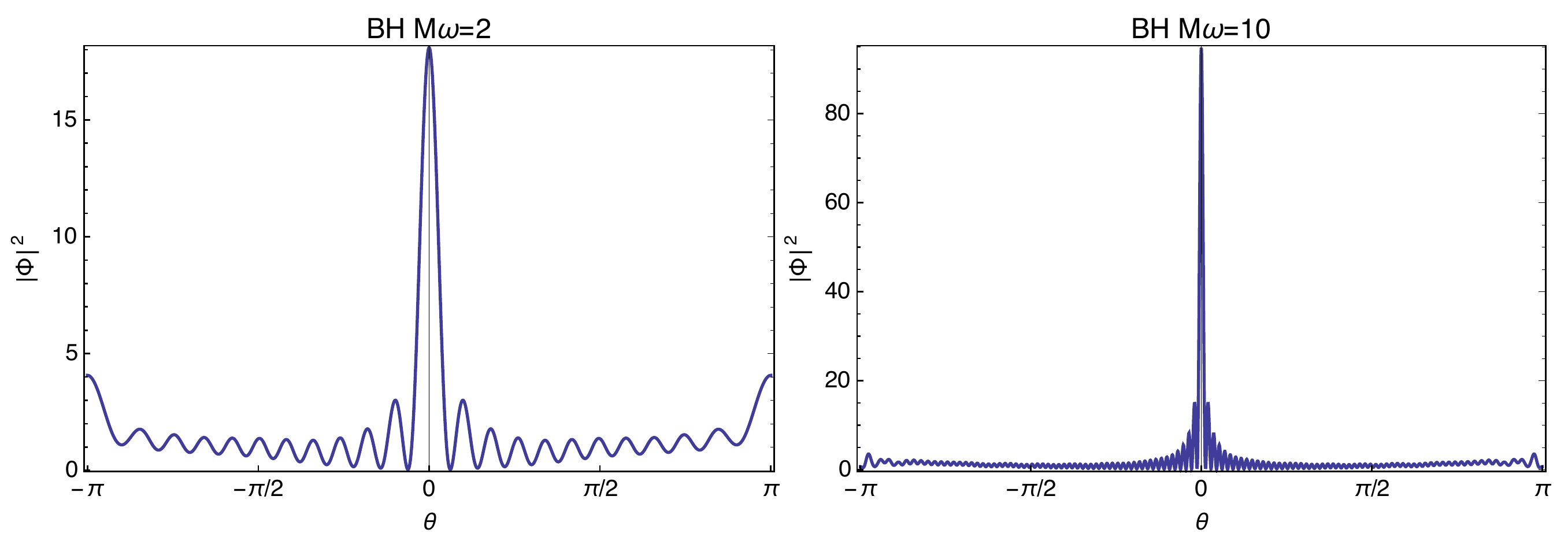}
    \caption{Sections of $M\omega=2$ and $M\omega=10$ of
      Fig.~\ref{fig:power-BH} (scattering amplitudes). These plots
      show interference fringes in the spatial domain.}
    \label{fig:power3-BH}
 \end{figure}
 \begin{figure}[H]
   \centering
   \includegraphics[width=1\linewidth,clip]{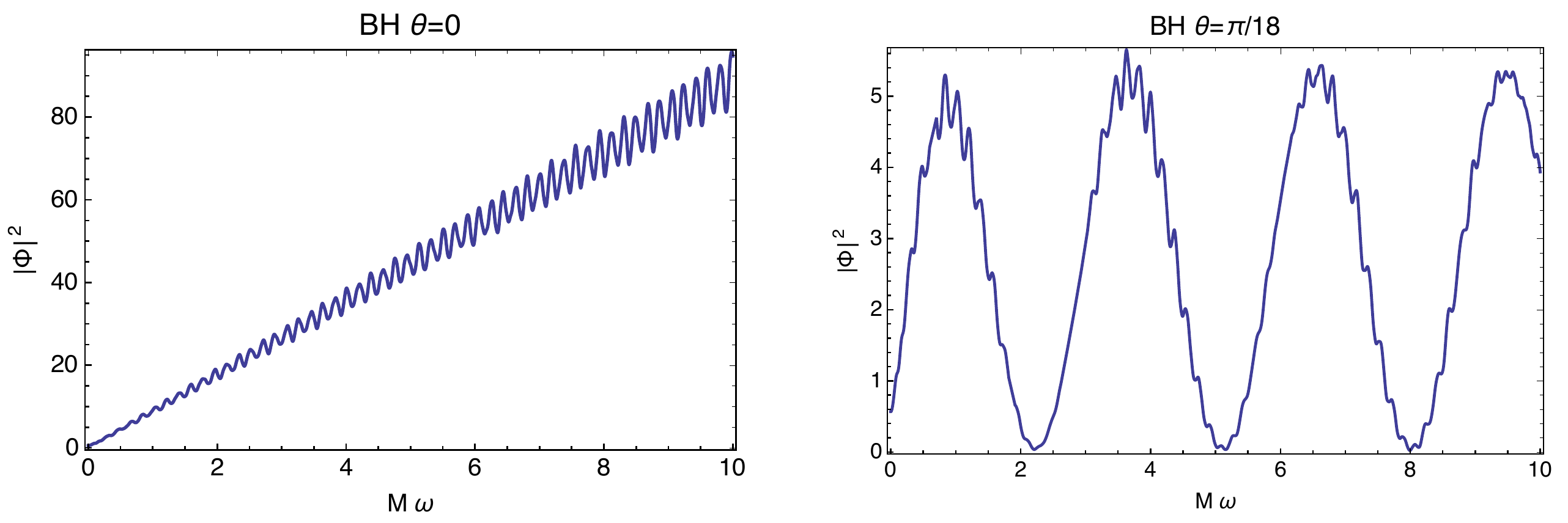}
   \caption{Sections of $\theta=0$ and $\theta=\pi/18$ of
     Fig.~\ref{fig:power-BH} (power spectrums). These plots show
     interference fringes in the frequency domain.}
   \label{fig:power2-BH} 
 \end{figure}

 We explain basic features of power spectrums
 (Fig.~\ref{fig:power2-BH}). We observe interference fringes in the
 power spectrums. For $\theta=\pi/18$, we have two components of
 oscillations. The component with the longer period is originated from
 interference between two light rays traveling far from the black hole
 (direct rays) and is associated with the weak gravitational lensing
 effect. \anno{The period of this oscillation is proportial to inverse
   of path difference ($\propto 1/\theta$) for small $\theta$ and for
   $\theta=0$ limit, the period becomes infinite and we do not have
   oscillation in the power spectrum caused by the interference
   between direct rays.}  The other component of oscillation has a
 shorter period $M\Delta\omega\sim 0.2$ which is independent of the
 scattering angle $\theta$ and exists even for the forward direction
 $\theta=0$; in this case the path difference between two direct rays
 becomes zero and we cannot expect interference fringe in the power
 spectrum. Thus we conclude that this oscillation of the power
 spectrum in the forward direction is caused by interference between
 winding rays and direct rays, and is peculiar to spacetimes with an
 unstable photon orbit (photon sphere). \anno{As another feature, the
   power spectrum at $\theta=0$ increases with $M\omega$. This is
   related to the caustics where the scattered waves are focused and
   the intensity diverges in the geometrical optics limit
   ($M\omega\rightarrow\infty$). The caustic points of a spherical
   lens are located right behind the lens object ($\theta=0$ line),
   and the sharpness of the divergence gets mild as the wavelength
   becomes large due to a wave effect. Therefore, we can see the
   feature in the left panel of Fig.~6.}

\begin{figure}[H]
   \centering
   \includegraphics[width=0.5\linewidth,clip]{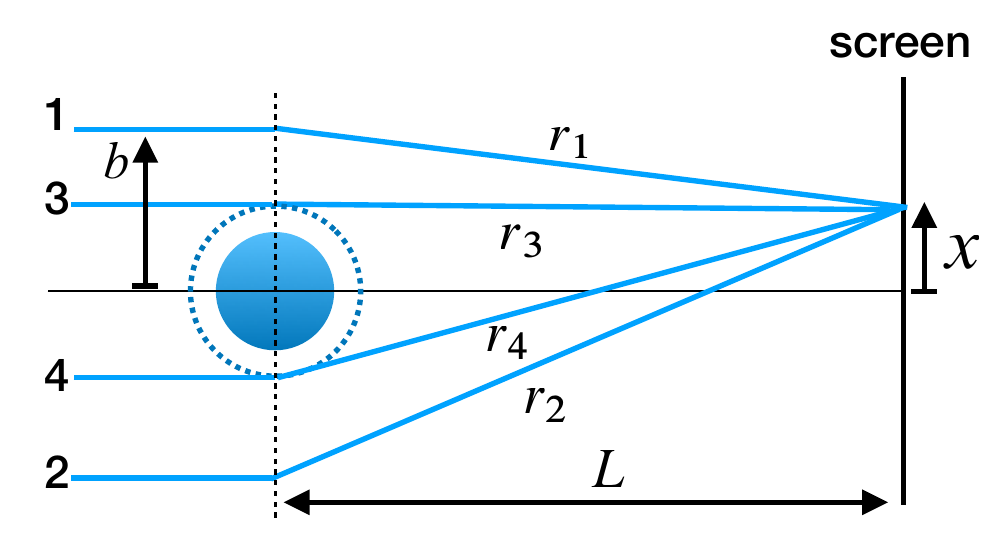}
   \caption{A model of gravitational lensing by a black hole. Ray 1
     and ray 2 represent direct rays of which the deflection angle is given
     by $\theta_\text{defl}$. Ray 3 and ray 4 represent winding rays.}
   \label{fig:toy}
 \end{figure}

 To clarify \anno{that} the period $M\Delta \omega\sim 0.2$
 corresponds to the scale of the unstable photon orbit, we consider a
 toy model of gravitational lensing by the black hole
 (Fig.~\ref{fig:toy}).
\noindent
We assume all rays follow straight lines as an approximation \anno{and
  the impact parameter $b$ is sufficiently smaller than $L$}. Rays 1
and 2 are direct rays and their deflection angle is assumed to obey
Einstein's formula
\begin{equation}
  \theta_\text{defl}=-\frac{4M}{b},
\end{equation}
where $b$ denotes the impact parameter of each ray. The position
\anno{$x$} on the screen  and $b$ are related by
\begin{equation}
  b_{1,2}=\frac{x\pm\sqrt{x^2+16LM}}{2}.
\end{equation}
Rays 3 and 4 correspond to winding rays with the impact parameter
$3\sqrt{3}\,M$. The path lengths of each ray are
\begin{align}
  & r_1=\sqrt{L^2+(b_1-x)^2},\quad r_2=\sqrt{L^2+(b_2-x)^2},\\
  & r_3=\sqrt{L^2+(b_3-x)^2}+2\pi\times |b_3|,\quad
    r_4=\sqrt{L^2+(b_4-x)^2}+2\pi\times |b_4|,
\end{align}
where $b_3=3\sqrt{3}M, b_4=-3\sqrt{3}M$. We assume winding rays go
around the black hole one round. Then ignoring the difference of
amplitudes for each ray, the wave on the screen is given by
\begin{equation}
  \Phi=e^{i\omega r_1}+e^{i\omega r_2}+c\left(e^{i\omega
      r_3}+e^{i\omega r_4}\right),
\end{equation}
where $c\approx 0.1$ represents the relative amplitude for winding rays
but the value of this constant does not affect the period of
interference in our estimation. For the forward direction
$\theta=0~\anno{(x=0)}$,
\anno{by neglecting $O(b/L)$ terms in the phase,}
we obtain
\begin{equation}
  |\Phi|^2\anno{/4}\approx 1+c^2+2c\cos\left[\omega\anno{(6\pi\sqrt{3}-2)}\,M\right],
\end{equation}
and the period of the power spectrum is given by
\begin{equation}
  M\Delta\omega=\frac{1}{\anno{3\sqrt{3}-1/\pi}}\approx\frac{1}{3\sqrt{3}}
    \approx 0.2.
\end{equation}
This value is consistent with the period of oscillation observed in
our numerical calculation.

\newpage
\subsection{Star case}  
\begin{figure}[H]  
  \centering
  \includegraphics[width=1\linewidth,clip]{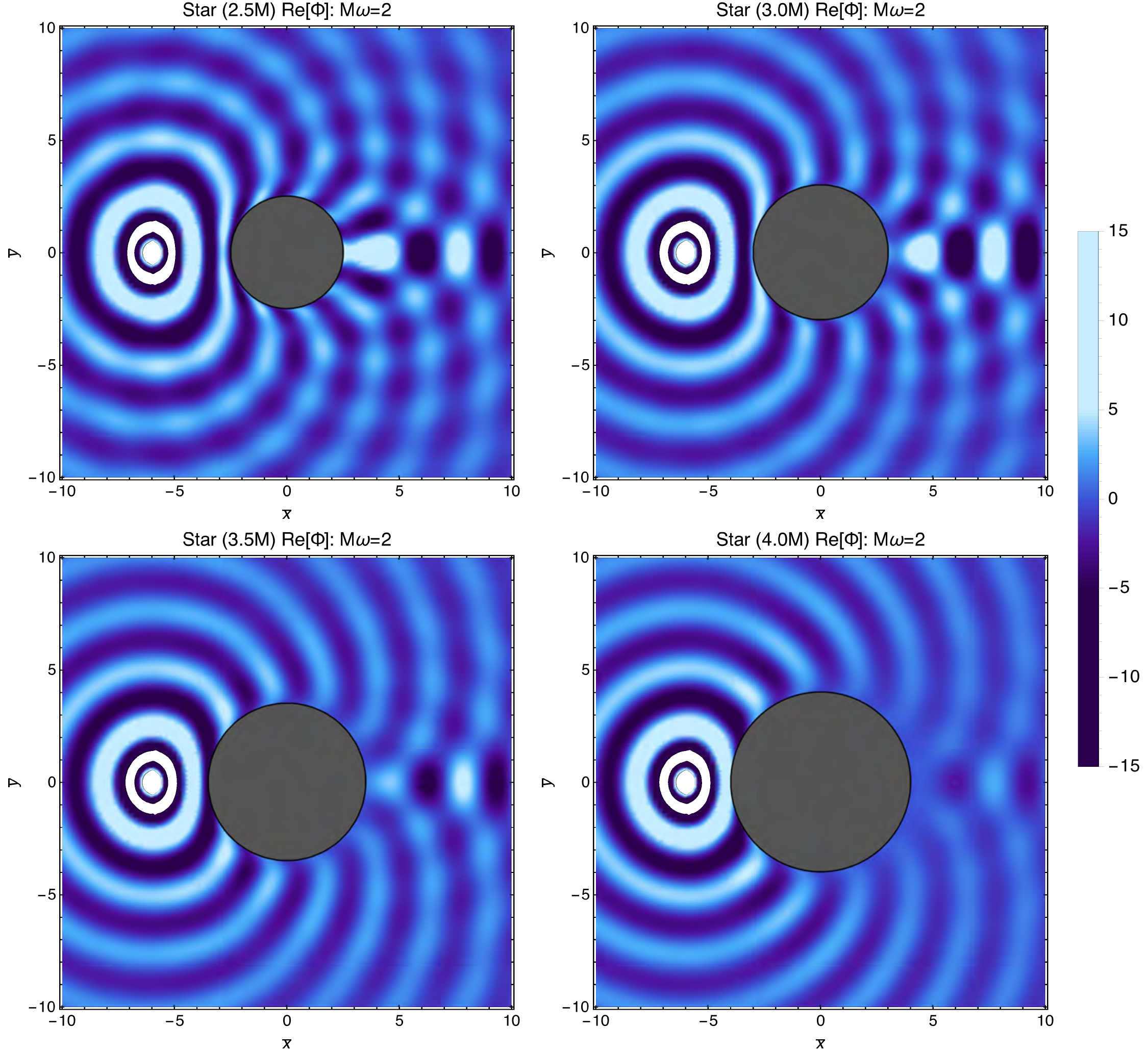}
  \caption{Real part of $\Phi$ with $M\omega=2$ for stars with a perfect
    absorbing surface. Four panels correspond to different values of the
    radii of stars $2.5M, 3M, 3.5M, 4M$.}
\label{fig:star-wavepattern}
\end{figure}

\begin{figure}[H]     
  \centering
  \includegraphics[width=1.\linewidth,clip]{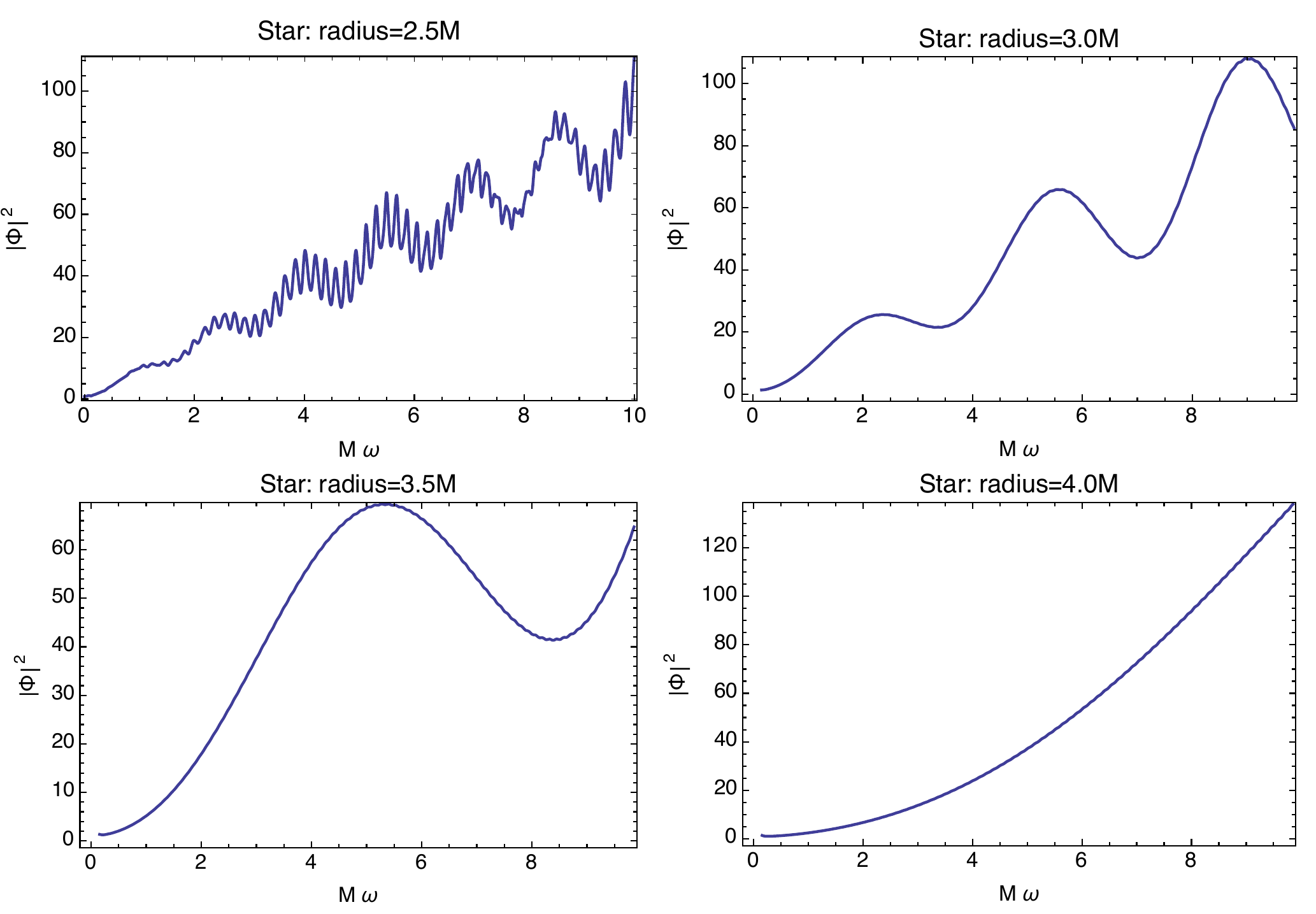}
  \caption{The power spectrum in the forward direction for stars
      with different
    radii  $2.5M, 3M, 3.5M, 4M$.}
  \label{fig:power-star} 
\end{figure}

To clarify wave effects associated with the photon sphere, 
we investigate stars with a perfect absorbing surface in the
Schwarzschild spacetime. We consider the following form of the effective
potential:
\begin{equation}
  V(r)=V_\text{BH}(r)\theta(r-r_\text{star}),
\end{equation}
where \anno{$\theta(r)$ is the unit step function, }$V_\text{BH}(r)$
denotes the effective potential \eqref{eq:pot-BH} of the Schwarzschild
spacetime and $r_\text{star}$ is the radius of the star. This form of
the potential models perfect absorptions of incoming waves at the
surface of the star:
\begin{equation}
  R_\ell(x_\text{tot})\propto 
e^{-i\omega x_\text{tot}}\quad \text{for}\quad r\leq r_\text{star}.
\end{equation}
We consider four different values of radii
$r_\text{star}=2.5M, 3M, 3.5M, 4M$. The obtained wave patterns in these models
are shown in Fig.~\ref{fig:star-wavepattern}. 
The star with radius $2.5M$, which is smaller than the photon sphere
of the Schwarzschild spacetime $3M$, is a model of black hole
mimickers (gravastar, boson star, etc.). For this case, the power
spectrum shows oscillation with two different periods (the upper left
panel in Fig.~\ref{fig:power-star}). The shorter one is
$M\Delta\omega_1\sim 0.2$, exactly the same value as the black hole case,
and is caused by interference between direct rays and winding rays
associated with the photon sphere. In addition, oscillations with
the longer period $M\Delta\omega_2\sim 2$ are superposed. We expect this
component is due to the diffraction effect caused by the surface of the
star. To justify this interpretation, we also investigated power
spectrums for stars
with other radii.
\noindent
For stars with radius larger than $3M$, the photon spheres are hidden
by the surface of stars and we do not have oscillation with
$M\Delta\omega_1\sim 0.2$. Power spectrums show oscillation with
the longer period $M\Delta\omega >2$ depending on the radius of the stars. It
is possible to estimate this period based on the diffraction effect of
waves (\anno{see} Sec.~IV).

\begin{figure}[H]
   \centering
    \includegraphics[width=1\linewidth,clip]{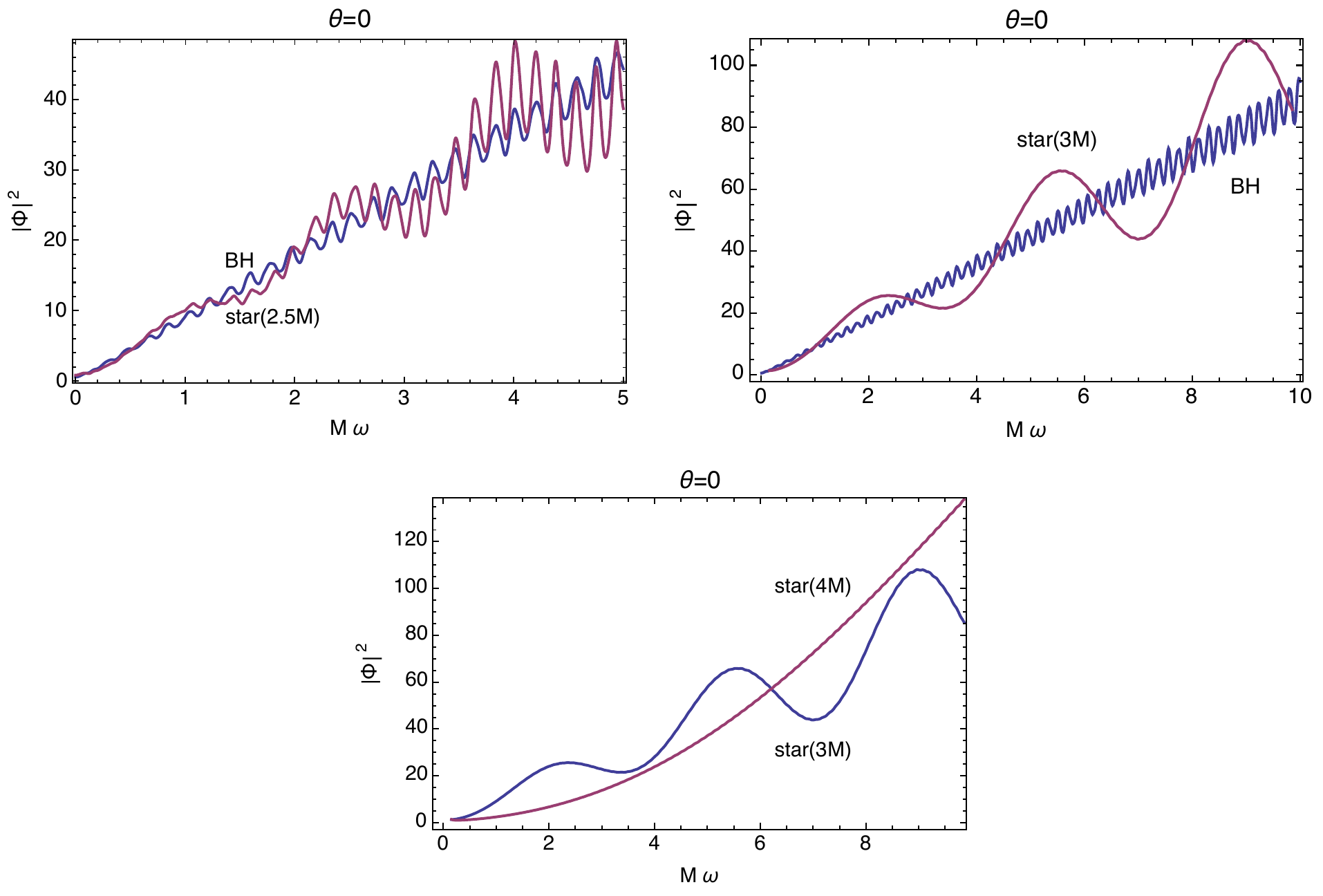}
  \caption{Power spectrum at $r_\text{obs}$ for the black hole and stars
    with radii $2.5M, 3M, 3.5M, 4M$. The first panel is plotted in
    the range $0\leq M\omega\leq 5$ to show the period of the
    oscillation clearly. }
  \label{fig:power-comp1}
\end{figure}

Figure~\ref{fig:power-comp1} summarizes the behavior of power spectrums
for the black hole case and stars with different radii cases.

\newpage 
\subsection{Wormhole case}  
We present the numerical result for the Ellis wormhole spacetime.
\paragraph{Massless case.} We choose parameters of the Ellis wormhole as
$m=0, ~a=3M$. In this case, the circumference radius of the wormhole
throat and the photon sphere coincide. The obtained wave patterns are
shown in Fig.~11.
\begin{figure}[H]
  \centering
  \includegraphics[width=1\linewidth,clip]{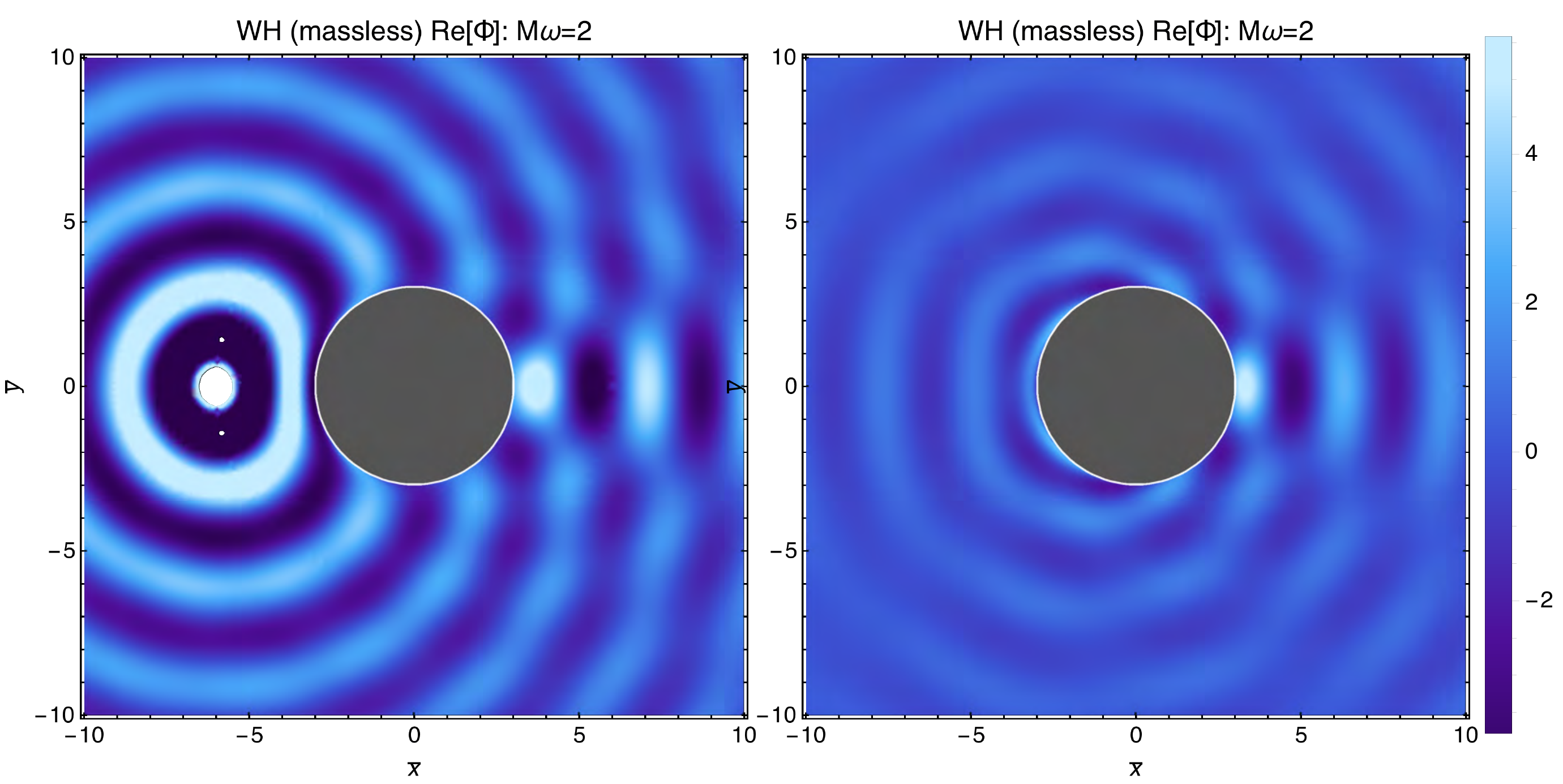}
  \caption{The real part of $\Phi$ for the massless wormhole with
    $M\omega=2$.  The left and right panels correspond to wormhole
    regions $x>0$ and $x<0$, respectively. The \anno{point wave}
    source is placed in $x>0$ region. The coordinates $\bar x$ and $\bar y$
    are introduced by
    $\bar{x}=\sqrt{x^2+a^2}\cos\theta,
    \bar{y}=\sqrt{x^2+a^2}\sin\theta$.}
\end{figure} 

The power spectrum has the same behavior as that of the black hole
(Fig.~12): it shows an oscillation with a period
$M\Delta\omega\sim 0.4$.
\begin{figure}[H]
  \centering  
  \includegraphics[width=1.\linewidth,clip]{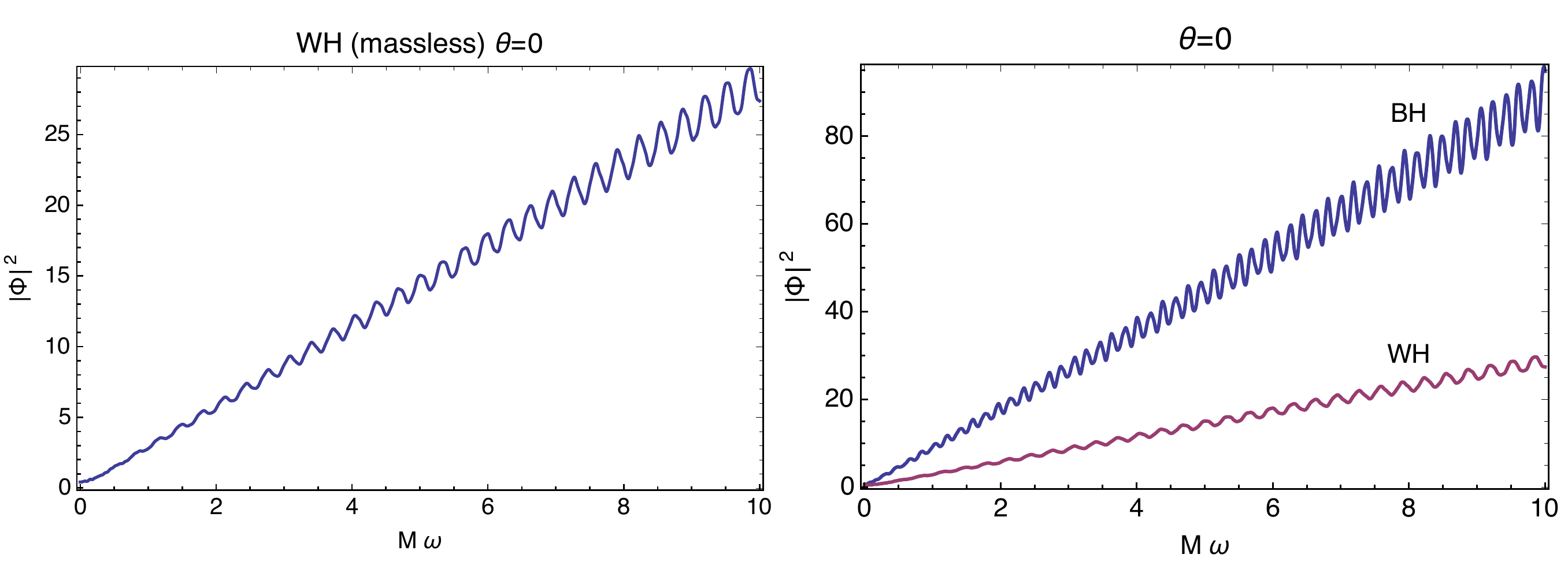}
  \caption{Power spectrum at $r_\text{obs}$ for the massless Ellis wormhole. }
\end{figure}
\noindent
We can explain this value using the same toy model presented in
Fig.~\ref{fig:toy}. For the massless Ellis wormhole, the deflection
angle is given by~\cite{1988}
\begin{equation}
  \theta_\text{defl}=\frac{\pi}{4}\left(\frac{a}{b}\right)^2.
\end{equation}
Then the period of oscillation in the power spectrum becomes
\begin{equation}
  M\Delta\omega=M\left[a-\frac{1}{4}\left(\frac{a}{2\pi
        L}\right)^{2/3}
    \right]^{-1}\sim 0.36
\end{equation}
for $a=3M$ and $L=20M$ and \anno{this formula} well explains the value
obtained by our numerical calculation. In this formula, the dependence
of $L$ (distance from the observer to the wormhole) appears due to
$b^{-2}$ behavior of the deflection angle, which is different from
that for the black hole and stars.

\paragraph{Massive case. } We choose the parameters of the wormhole as
$m=M,~a=1.305716M$. For these parameters, the size of the throat is
$3M$ and the photon sphere is $3.4823M$.

\begin{figure}[H]
  \centering
  \includegraphics[width=1\linewidth,clip]{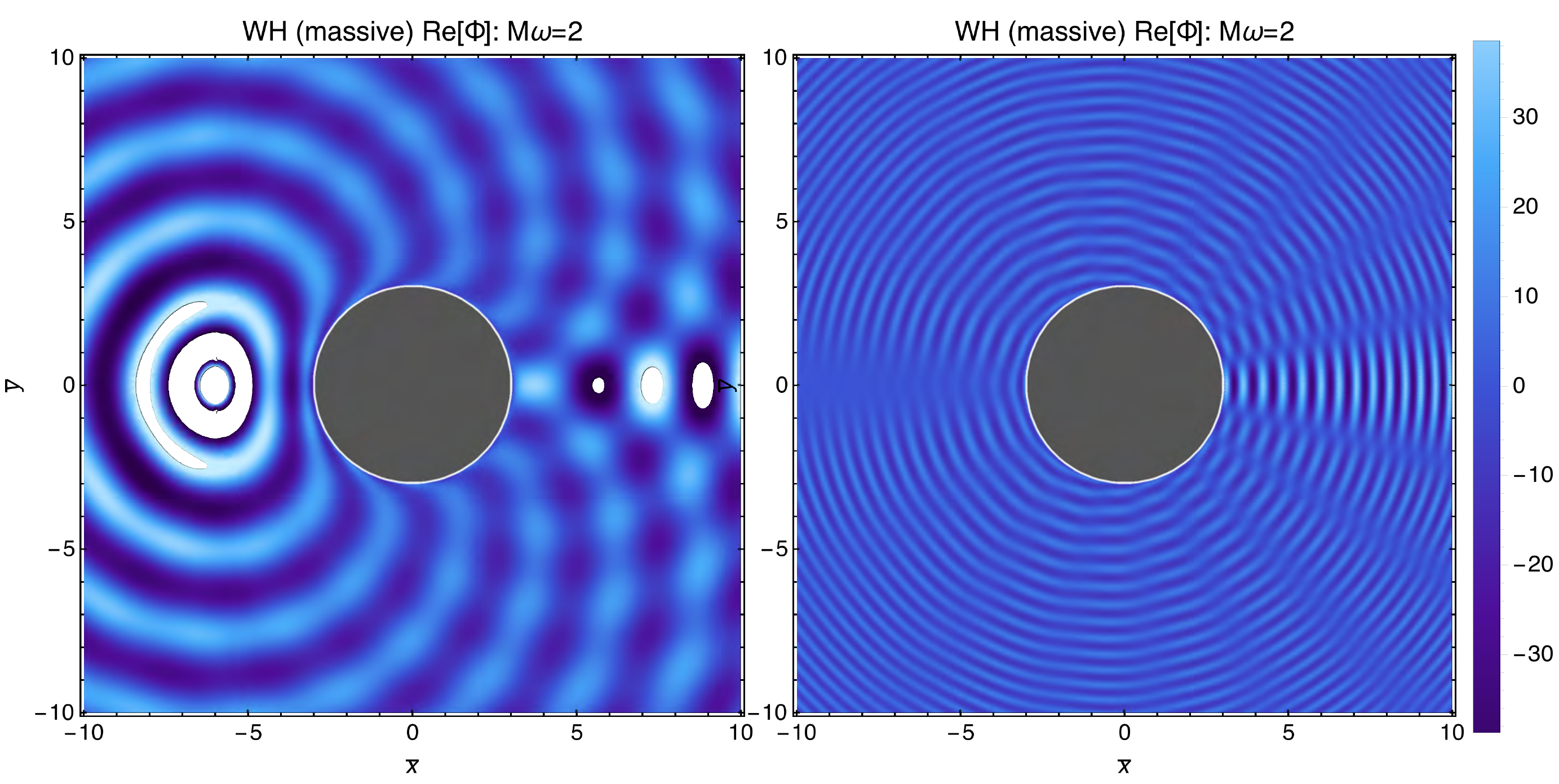}
  \caption{The real part of $\Phi$ for the massive wormhole with
    $M\omega=2$.  The left and right panels correspond to wormhole
    regions $x>0$ and $x<0$, respectively. The coordinates $\bar x$
    and $\bar y$ are
    introduced as
    $\bar{x}=\sqrt{x^2+a^2-m^2}\cos\theta,
    \bar{y}=\sqrt{x^2+a^2-m^2}\sin\theta$.}
\end{figure} 
We notice that intervals of the wave front are different for the $x>$
region and the $x<0$ region (Fig.~13). This is caused by different
asymptotic behaviors of metric \eqref{eq:metric-wh}. For
$x\rightarrow \infty$, the metric is
\begin{equation}
  ds^2\approx
  -\left(1-\frac{2m}{r}\right)dt^2+\left(1+\frac{2m}{r}\right)dr^2
  +r^2 d\Omega^2,
\end{equation}
whereas for $x\rightarrow-\infty$,
\begin{equation}
  ds^2\approx -\al^{-2}\left(1+\frac{2\al
      m}{r}\right)dt^2+\left(1-\frac{2\al m}{r}\right)dr^2+r^2 d\Omega^2,
\end{equation}
with $\al=\exp\left(\pi m/\sqrt{a^2-m^2}\right)$ and this \anno{metric
  represents}  a spacetime
with negative gravitational mass $-\al  m$. Let us consider the
radial null vector $k_\mu=(\omega, k)$. Then $\omega$ and $k$ are connected
by the relation
\begin{equation}
  k=\omega\sqrt{\left|\frac{g^{tt}}{g^{rr}}\right|}=
  \begin{cases}
    \omega(1+2m/r)&\quad\text{for}\quad x\rightarrow\infty\\
    \omega\al(1-2\al m/r)&\quad\text{for}\quad x\rightarrow-\infty
  \end{cases}
\end{equation}
As the interval of wave front $\Delta r$ is determined by
$k\Delta r=\text{const}$, thus $\Delta r\propto 1/k$. For
$\al\gg 1$, which holds for values of present parameters, the interval
of the wave front in the region $x<0$ becomes much smaller compared to
that in the region $x>0$.

\begin{figure}[H]
  \centering  
  \includegraphics[width=1.\linewidth,clip]{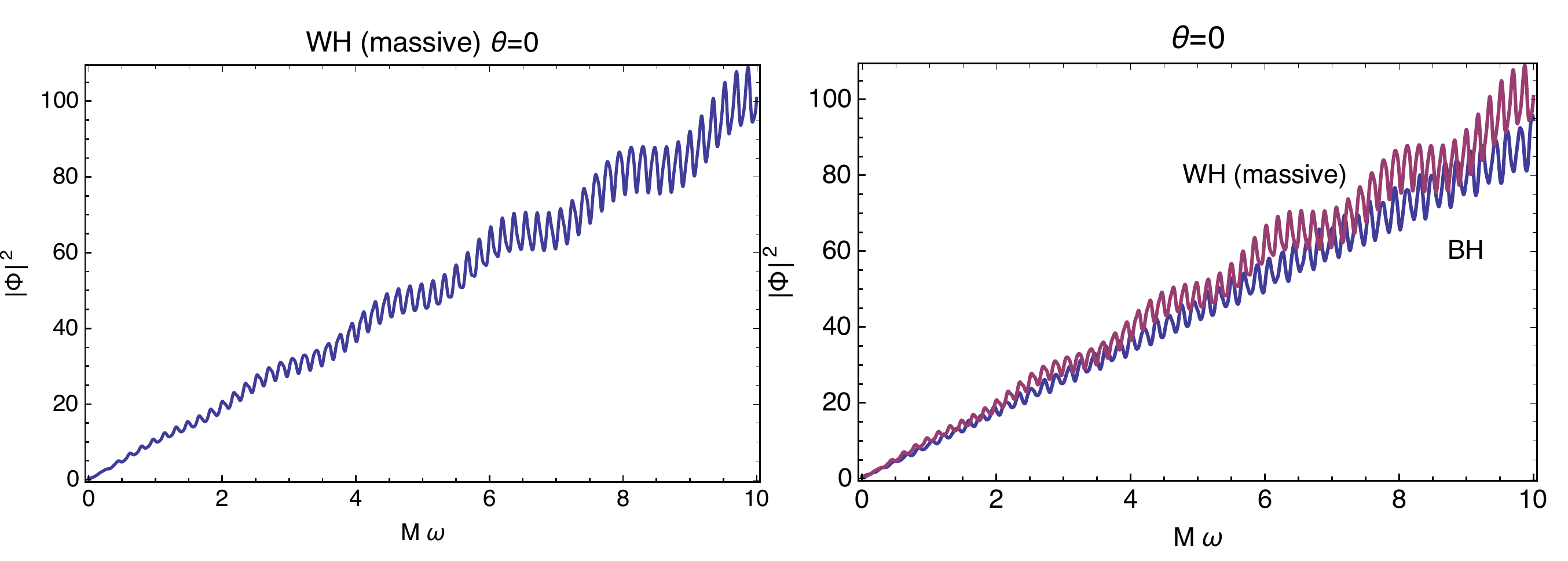}
  \caption{Left: Power spectrums at $r_\text{obs}$ for the massive Ellis
    wormhole. Right: Power spectrums for the black hole and the
    massive Ellis wormhole.}
  \label{fig:power-wh2}
\end{figure}

Figure~\ref{fig:power-wh2} shows the power spectrum for the massive wormhole.
\noindent
Power spectrums have oscillations with two different periods.  The
shorter one $M\Delta\omega_1\sim 0.2$ is due to interference between
direct rays and winding rays, and the value coincides with that for
the black hole because the radius of the photon sphere and the
deflection angle for direct rays are the same as the black hole.  The
longer one $M\Delta\omega_2\sim 2$ comes from the diffraction effect
by the absorbing region: for the massive wormhole, this region
corresponds to the wormhole throat $3M$ which is smaller than the
radius of photon sphere $3.48M$.

\newpage
\section{Interpretation of power spectrum oscillations}
In this section, we will provide theoretical justification for
behaviors of the power spectrums based on analytic formulas of wave
optics.  We have two key factors associated with wave effects in our
scattering problem: interference and diffraction.

\begin{figure}[H] 
  \centering
  \includegraphics[width=1\linewidth,clip]{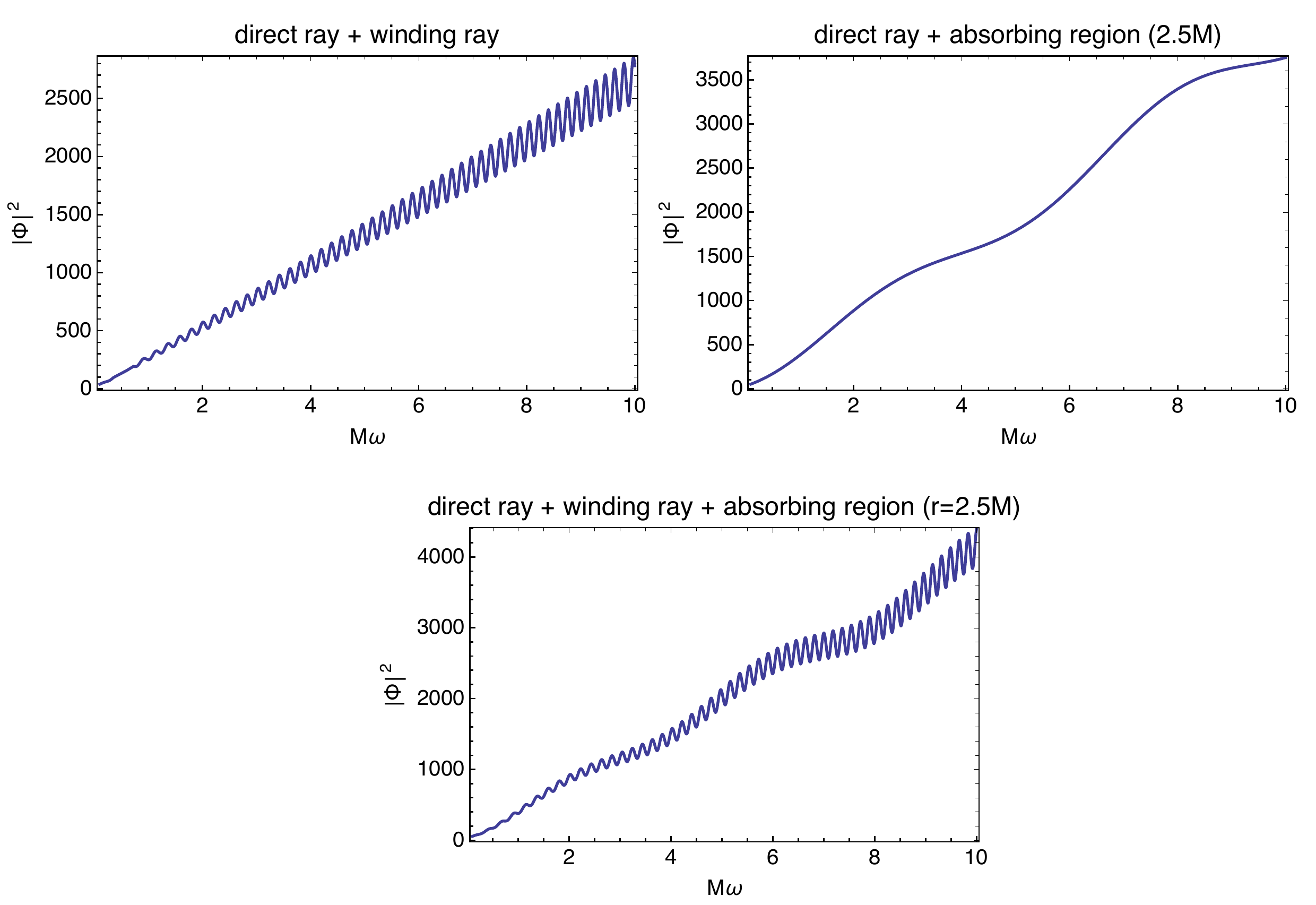}
  \caption{Effect of interference and diffraction in  power
    spectrums. The first panel is the power spectrum obtained by taking
    into account the interference between  direct rays and 
    winding rays. The second panel is the power spectrum obtained by
    taking into account the diffraction effect for  direct
    rays. The third panel is the power spectrum obtained by taking into
    account  all effects. We assume $b_0=2.5M$ and $b_c=3\sqrt{3}M$ in
    \anno{these plots}.}
  \label{fig:pow-ana}
\end{figure}

At a distant observing point $r$, the scattering wave from a point
\anno{wave} source at $r_s$ is represented as \cite{Nambu2016}
\begin{equation}
  \Phi(r,\theta)\approx\frac{e^{i\omega(x+x_s)}}{4\pi
    i\omega rr_s}\sum_{\ell=0}^\infty
  \lambda\,e^{i\frac{\lambda^2}{2\omega
      \bar{r}}}\,e^{2i\del_\ell}P_\ell(\cos\theta),\quad
    \lambda:=\ell+\frac{1}{2},
\end{equation}
where $x$ and $x_s$ are tortoise coordinates corresponding to $r$ and
$r_s$, and $\bar{r}=r r_s/(r+r_s)$. $\del_\ell$ represents the phase
shift. Applying Poisson's sum formula, we can replace the sum with
respect to $\ell$ to the integral over continuous variable
$\lambda$. In the eikonal limit, it can be shown that
\begin{equation}
  \Phi(r,\theta)\approx \frac{e^{i\omega(x+x_s)}}{4\pi i \omega r r_s}
\left[\int_0^\infty
  d\lambda\,\lambda\,e^{i\frac{\lambda^2}{2\omega\bar{r}}}\,
  e^{2i\del_{\lambda-1/2}}\,J_0(\lambda\theta)+2\pi
  i\sum_{n=0}^\infty\lambda_n\,\gamma_n\,e^{i\frac{\lambda_n^2}{2\omega\bar{r}}}
  J_0(\lambda_n\theta)f(\lambda_n)\right],
  \label{eq:scat-wave}
 \end{equation}
where \anno{$J_0$ is the Bessel function with the 0th order and}
\begin{align}
  &\lambda_n=\lambda_c+i\left(n+\frac{1}{2}\right),\quad
    \lambda_c=3\sqrt{3}M\omega, \\
  &\gamma_n=-\frac{i}{\sqrt{2\pi}}\left(\frac{\lambda_n}{\lambda_c}\right)
    \left(n+\frac{1}{2}\right)^{n+1/2}\,\frac{e^{-(n+1/2)}}{n!},\quad
f(\lambda_n)=\frac{1}{1-e^{-2i\pi(\lambda_n-1/2)}}.
\end{align}
The first term in \eqref{eq:scat-wave} corresponds to the
Fresnel-Kirchhoff diffraction formula in the wave
optics~\cite{Sharma2006}, 
and the second term comes from the contribution of poles in the $S$ matrix
$e^{2i\del_\ell}$ in the complex $\ell$ plane (Regge poles).  This
term represents the orbiting effect associated with the photon
sphere. After taking the $n$ sum, we obtain~\cite{Nambu2016}
\begin{equation}
  \Phi(r,\theta)\propto \frac{1}{\omega}\int_0^\infty
  d\lambda\,\lambda\,e^{i\frac{\lambda^2}{2\omega\bar{r}}}\,
  e^{2i\del_{\lambda-1/2}}\,J_0(\lambda\theta)+\frac{1}{\omega}
  \sqrt{\frac{\pi}{2}}\,e^{-\pi-i\pi/4+i\pi\lambda_c}\,\lambda_c\,e^{\frac{i\lambda_c^2}{2\omega\bar{r}}}\,\sqrt{\omega\bar{r}}\,J_0(\lambda_c\theta).
\end{equation}
For the forward direction $\theta=0$,
\begin{align}
  \Phi &\propto \frac{1}{\omega}\int_0^\infty
  d\lambda\,\lambda\,e^{i\frac{\lambda^2}{2\omega\bar{r}}}\,
  e^{2i\del_{\lambda-1/2}}+\frac{1}{\omega}
  \sqrt{\frac{\pi}{2}}\,e^{-\pi-i\pi/4+i\pi\lambda_c}\,\lambda_c\,e^{\frac{i\lambda_c^2}{2\omega\bar{r}}}\,\sqrt{\omega\bar{r}}
                  \notag \\
  &=\omega\int_{b_0}^\infty
  db\,b\,e^{i\frac{\omega b^2}{2\bar{r}}}\,
  e^{2i\del_b}+
  \sqrt{\frac{\pi}{2}}\,e^{-\pi-i\pi/4+i\pi\omega b_c}\,b_c
    \,e^{\frac{i\omega b_c^2}{2\bar{r}}}\,\sqrt{\omega\bar{r}},
    \label{eq:scat-wave2}
\end{align}
where we introduced the impact parameter $b=\lambda/\omega$ and a
lower cutoff $b_0$ of the integral, which represents the effect of a
\anno{perfect} absorbing region (the black hole horizon, surface of the
stars). Concerning the form of the phase shift for direct rays, we
adopt $\del_b=-2M\omega\ln(b\omega)$ which results in the scattering
angle in the eikonal limit as
\begin{equation}
  2\frac{d}{d(b\omega)}\del_b=-\frac{4M}{b},
\end{equation}
and reproduces Einstein's formula of deflection angle.
Although this formula is correct only for rays with sufficiently
large impact parameters compared to the size of the photon sphere, it is
adequate for our purpose here to obtain a qualitative understanding of
the oscillation of power spectrums.  Then after performing the integral,
the first term in Eq.~\eqref{eq:scat-wave2} becomes
\begin{equation}
  \bar{r}\left(2i\bar{r}\omega\right)^{-2iM\omega}\left[
  2M\omega\Gamma(-2iM\omega)-i\left\{\Gamma(1-2iM\omega)-\Gamma\left(
1-2iM\omega, -\frac{ib_0^2\omega}{2\bar{r}}\right)\right\}\right],
\label{eq:scat-wave3}
\end{equation}
where the third term denotes the incomplete gamma
function. We show the behavior of the obtained analytic formula in
Fig.~\ref{fig:pow-ana}. 
\noindent
The analytic formula \eqref{eq:scat-wave2} well reproduces behaviors
of the power spectrum obtained by our numerical calculation. We can
estimate the period of oscillations in the power spectrum. Using
\eqref{eq:scat-wave2} and \eqref{eq:scat-wave3}, the period of
oscillation \anno{due to the}  diffraction effect is
\begin{equation}
  M\Delta\omega_2\sim
  2\pi\left(2-\frac{b_0^2}{2M\bar{r}}\right)^{-1}\sim 4.8
\quad \text{(for $b_0=2.5M,
\bar{r}=4.6M$)}.
\end{equation}
Although this value is about 2 times larger than the value obtained by
the numerical calculation, the formula shows an increase of period for
the larger size of the diffraction region and qualitatively explains
the behavior of the power spectrum obtained by numerical calculations.
The period of the oscillation due to interference between direct rays
and winding rays is
\begin{equation}
  M\Delta\omega_1\sim \frac{M}{b_c-M/\pi}\sim 0.2\quad 
\text{(for $b_c=3\sqrt{3}\,M$)}. 
\end{equation}
These values are consistent with the period of oscillations in the
power spectrums obtained by the numerical calculation for the black
hole and stars with smaller radius than the photon sphere.
\section{Summary and conclusion}

By solving the scalar wave equation numerically, we obtained the
scattering wave by the Schwarzschild black hole, the spherical star with a
perfect absorbing surface, and the Ellis wormhole, and then we
investigated the wave pattern and the power spectrums.  We focused on
the case that an observer is located at the forward position where we
\anno{do not} expect the interference between direct rays due to the
difference of paths lengths in the geometrical optics point of view.
Even in this case, we found two kinds of oscillations in the power
spectrums.  When a gravitating object has the photon sphere (black
hole, star with $r_\text{star}\leq 3M$ and Ellis wormholes), we can
see the oscillation of the power spectrum reflecting the interference
between the direct ray and the winding ray.  Moreover, diffraction effects
\anno{due to the absorption boundary condition} were observed for
stars and massive Ellis wormholes .  We have justified the periods of
these oscillations by analytic evaluation of the wave scattering.  As
expected, it is possible to distinguish black holes from their mimickers
by looking inside of the photon sphere \anno{using} waves although we cannot
tell the difference in the geometrical optics for the present
source-object-observer configuration.

As other interesting models, we will consider stars with an internal
structure or a reflecting surface and wormholes with a double-peak
effective potential \cite{Bueno2018}, which may give echoes in the power
spectrums of the scattered waves.

%

\acknowledgments{
Y. N. was supported in part by JSPS KAKENHI Grant Number
15K05073 and  S. N. was supported
by JSPS KAKENHI Grant No. 17J10770.}

\appendix
\section{Wormhole spacetimes}
As an example of spacetimes with the photon unstable orbit without
horizon, we consider the Ellis wormhole~\cite{Ellis1973}.  Although
the wormhole spacetimes are unstable and may not \anno{be} realized in
our universe, we use them as benchmark models to detect wave optical
effects for compact gravitating objects.

The Ellis wormhole spacetime is obtained as the solution of the
Einstein-scalar system
\begin{equation}
  R_{\mu\nu}=2\chi_{,\mu}\chi_{,\nu},\quad \square\chi=0.
\end{equation}
The metric  is given by~\cite{Ellis1973}
\begin{align}
  &ds^2=-f dt^2+\frac{1}{f}\left(dx^2+(x^2+a^2-m^2)d\Omega^2\right),\quad
    -\infty<x<+\infty, \label{eq:metric-wh}\\
  &f=\exp\left(-\frac{2m\chi(x)}{a}\right),\quad\chi(x)=\frac{a}{\sqrt{a^2-m^2}}
    \left(\frac{\pi}{2}-\mathrm{arctan}\frac{x}{\sqrt{a^2-m^2}}\right),
\end{align}
where $a$ and $m$ are constants representing the throat size 
 and the mass of the wormhole, respectively. 
The range of the scalar field is $0\,(x=+\infty)\le\chi\le\frac{\pi
  a}{\sqrt{a^2-m^2}}\,(x=-\infty)$. By introducing a new radial
coordinate corresponding to the circumference radius
\begin{equation}
  r=\frac{\sqrt{x^2+a^2-m^2}}{f^{1/2}},
\end{equation}
the metric becomes
\begin{equation}
  ds^2=-fdt^2+\frac{dr^2}{\anno{h}}+r^2d\Omega^2,\quad \anno{h}=f\left(\frac{dr}{dx}\right)^2.
  \label{eq:wh-metric}
\end{equation}
In terms of the scalar field $\chi$,
\begin{align}
  &x=\sqrt{a^2-m^2}\,\cot\left(\frac{\sqrt{a^2-m^2}}{a}\chi\right),\\
  &r=\frac{\sqrt{a^2-m^2}\,e^{m\chi/a}}{\sin\left(\frac{\sqrt{a^2-m^2}}{a}
\chi\right)},\quad r_{\text{min}}=a\exp\left[\frac{m}{\sqrt{a^2-m^2}}\mathrm{Arccot}\left(\frac{m}{\sqrt{a^2-m^2}}\right)\right],\\
  &\anno{h}=\left[\cos\left(\frac{\sqrt{a^2-m^2}}{a}\chi\right)-\frac{m}{\sqrt{a^2-m^2}}\sin\left(\frac{\sqrt{a^2-m^2}}{a}\chi\right)
\right]^2.
\end{align}
$r_\text{min}$ is the circumference radius of the wormhole throat (at
$r=m$).  The asymptotic behavior of the metric \eqref{eq:wh-metric} for
$x\rightarrow\infty$ ($\chi\rightarrow 0$) is
\begin{align*}
  &x\sim \frac{a}{\chi},\quad R\sim\frac{a}{\chi}\sim x,\\
  &\anno{h}\sim
   1-\frac{2m}{x}-\frac{a^2-2m^2}{x^2},\quad
  f\sim 1-\frac{2m\chi}{a}\sim1-\frac{2m}{x}.
\end{align*}
Thus the metric for $x\rightarrow\infty$ becomes
\begin{equation}
  ds^2\approx-\left(1-\frac{2m}{r}\right)dt^2+\left(1+\frac{2m}{r}
+\frac{a^2+2m^2}{r^2}\right)
  dr^2+r^2 d\Omega^2.
\end{equation}
On the other hand, the asymptotic behavior of the metric for
$x\rightarrow-\infty$ ($\chi\rightarrow\frac{\pi a}{\sqrt{a^2-m^2}}$)
is
\begin{align*}
  &x\sim \frac{-1}{\frac{\pi}{\sqrt{a^2-m^2}}-\frac{\chi}{a}},\quad 
r\sim -x\, e^{\frac{\pi m}{\sqrt{a^2-m^2}}},\quad \anno{h}\sim
  1-\frac{2m}{x}\sim 1+\frac{2m\, e^{\frac{\pi
    m}{\sqrt{a^2-m^2}}}}{r},\\
  &f\sim e^{-\frac{2\pi
    m}{\sqrt{a^2-m^2}}}\left(1-\frac{2m}{x}\right)
    \sim e^{-\frac{2\pi
    m}{\sqrt{a^2-m^2}}}\left(1+\frac{2m\, e^{\frac{\pi
    m}{\sqrt{a^2-m^2}}}}{r}\right).
\end{align*}
Thus the metric for $x\rightarrow -\infty$ represents a gravitating
object with negative mass $-m\, e^{\frac{\pi m}{\sqrt{a^2-m^2}}}$. For
massless case $m=0$, the metric reduces to
\begin{equation}
  ds^2=-dt^2+dx^2+(x^2+a^2)d\Omega^2.
\end{equation}
In this case, the gravitational potential in the weak field region
behaves as $\propto r^{-2}$ and the law of gravity is different from
that for a point mass.

\begin{figure}[H]
  \centering
  \includegraphics[width=0.5\linewidth,clip]{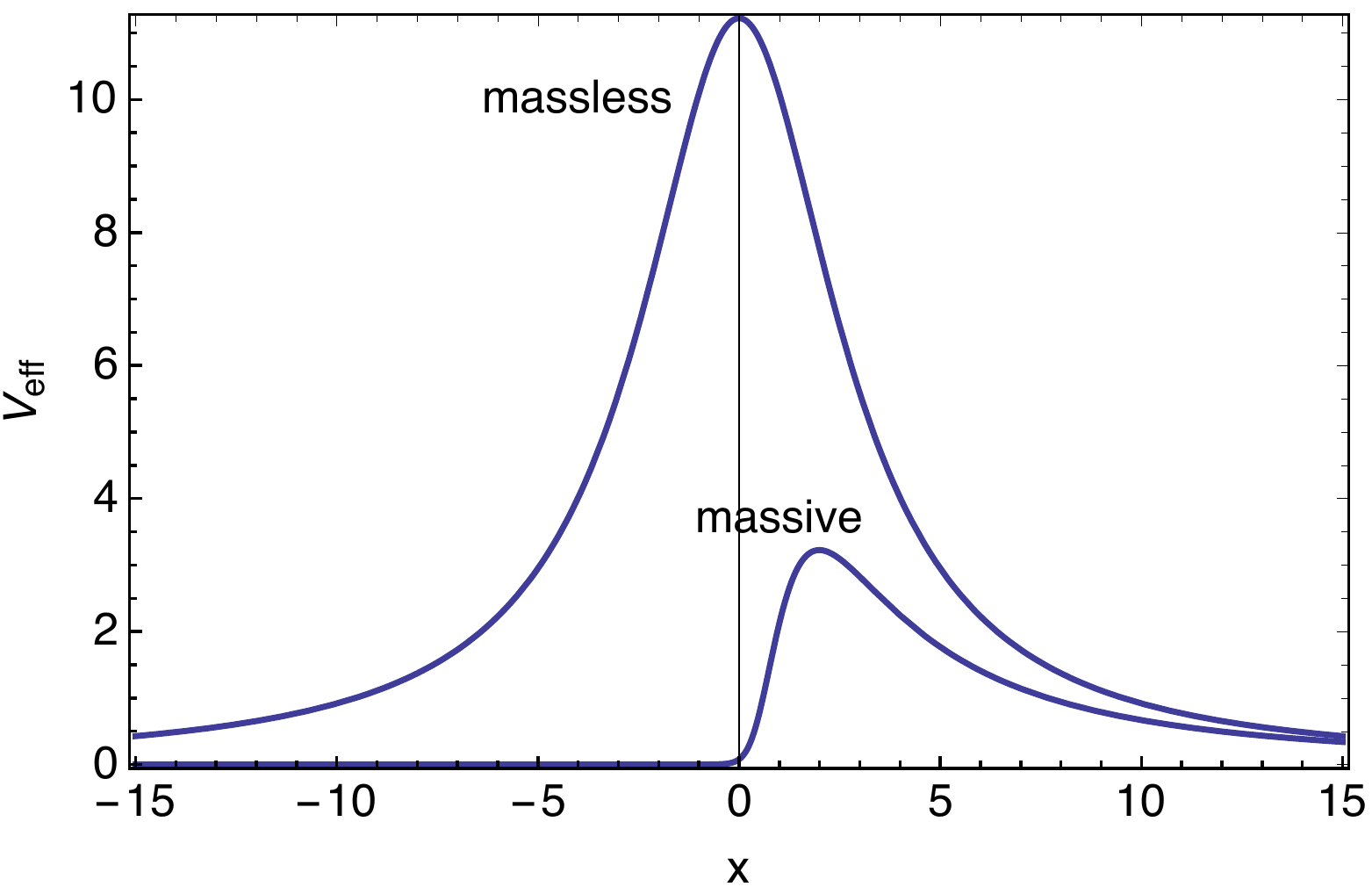} 
  \caption{Effective potentials for the Ellis wormhole ($\ell=10$) for
    the massless case ($a=3M, m=0$) and the massive case
    ($a=1.305716M, m=M$). The throat of the wormhole is located at
    $x=m$ and the peak of the potential in the eikonal limit
    $\ell\gg 1$ is at $x\approx 2m$. This place corresponds to the
    photon sphere. }
  \label{fig:pot-wh}
\end{figure}

The shape of the effective potential for the wormhole
is shown in Fig.~\ref{fig:pot-wh}.
\noindent
In our numerical calculations, we adopt the wormhole parameters as
$a=3M, m=0$ (massless case) and $a=1.305716M, m=M$ (massive case). 
The circumference radii of the throat for both wormholes are $3M$ and the
circumference radii of the photo sphere are $3M$ (massless case) and
$3.4823M$ (massive case).


\begin{thebibliography}{10}
\newcommand{\enquote}[1]{``#1''}

\bibitem{EHT}
T.~E. H.~T. Collaboration, \enquote{First M87 Event Horizon Telescope Results.
  I. The Shadow of the Supermassive Black Hole}, \emph{The Astrophys. 
  J.  Lett.} \textbf{875}, (2019) 875:L1.

\bibitem{Cunha2017}
P.~V.~P. Cunha, C.~A.~R. Herdeiro, and E.~Radu, \enquote{{Fundamental photon
  orbits : Black hole shadows and spacetime instabilities}}, \emph{Phys. Rev.
  D} \textbf{96}, (2017) 024039.

\bibitem{Cunha2017a}
P.~V.~P. Cunha, A.~Font, C.~Herdeiro, E.~Radu, N.~Sanchis-gual, and
  M.~Zilh{\~{a}}o, \enquote{{Lensing and dynamics of ultracompact bosonic
  stars}}, \emph{Phys. Rev. D} \textbf{96}, (2017) 104040.

\bibitem{Cunha2017b}
P.~V.~P. Cunha, E.~Berti, and C.~A.~R. Herdeiro, \enquote{{Light-Ring Stability
  for Ultracompact Objects}}, \emph{Phys. Rev. Lett.} \textbf{119}, (2017)
  251102.

\bibitem{Cunha2018}
P.~V.~P. Cunha, C.~A.~R. Herdeiro, and P.~V.~P. Cunha, \enquote{{Shadows and
  strong gravitational lensing : a brief review}}, \emph{Gen. Relativ.
Gravit.} \textbf{50}, (2018) 1--27.

\bibitem{Cunha2018a}
P.~V.~P. Cunha, C.~A.~R. Herdeiro, and M.~J. Rodriguez, \enquote{{Shadows of
  exact binary black holes}}, \emph{Phys. Rev. D} \textbf{98}, (2018)
  44053.

\bibitem{Cunha2018b}
P.~V.~P. Cunha, C.~A.~R. Herdeiro, and M.~J. Rodriguez, \enquote{{Does the
  black hole shadow probe the event horizon geometry ?}}, \emph{Phys. Rev. D}
  \textbf{97}, (2018) 84020.

\bibitem{Sakai2014}
N.~Sakai, \enquote{{Gravastar shadows}}, \emph{Phys. Rev. D} \textbf{90},
  (2014) 104013.

\bibitem{Ohgami2015}
T.~Ohgami and N.~Sakai, \enquote{{Wormhole shadows}}, \emph{Phys. Rev. D}
  \textbf{91}, (2015) 124020.

\bibitem{Ohgami2016}
T.~Ohgami and N.~Sakai, \enquote{{Wormhole shadows in rotating dust}},
  \emph{Phys. Rev. D} \textbf{94}, (2016) 064071.

\bibitem{Cardoso2009}
V.~Cardoso, A.~Miranda, E.~Berti, H.~Witek, and V.~Zanchin, \enquote{{Geodesic
  stability, Lyapunov exponents, and quasinormal modes}}, \emph{Phys. Rev. D}
  \textbf{79}, (2009) 064016.

\bibitem{Andersson2000}
N.~Andersson and B.~Jensen, \enquote{{Scattering by black holes}}, \emph{arXiv:gr-qc/0011025}.

\bibitem{Glampedakis2001}
K.~Glampedakis and N.~Andersson, \enquote{{Scattering of scalar waves by
  rotating black holes}}, \emph{Class. Quantum Grav.} \textbf{18}, (2001)
  1939--1966.


\bibitem{Kanai2013}
K.-i. Kanai and Y.~Nambu, \enquote{{Viewing black holes by waves}},
  \emph{Class. Quantum Grav.} \textbf{30}, (2013) 175002.

\bibitem{Nambu2016}
Y.~Nambu and S.~Noda, \enquote{{Wave optics in black hole spacetimes: the
  Schwarzschild case}}, \emph{Class. Quantum Grav.} \textbf{33}, (2016)
  075011.

\bibitem{Matzner1968}
R.~A. Matzner, \enquote{{Scattering of massless scalar waves by a Schwarzschild
  " singularity"}}, \emph{J. Math. Phys.} \textbf{9},
  (1968) 163--170.

\bibitem{Sanchez1978}
N.~S\'anchez, \enquote{Elastic scattering of waves by a black hole},
  \emph{Phys. Rev. D} \textbf{18}, (1978) 1798--1804.

\bibitem{Handler1980}
F.~A. Handler and R.~A. Matzner, \enquote{{Gravitational wave scattering}},
  \emph{Phys. Rev. D} \textbf{22}, (1980) 2331--2348.

\bibitem{Zhang1984}
T.-r. Zhang and C.~Dewitt-morette, \enquote{{WKB Cross Section for Polarized
  Glories of Massless Waves in Curved Space-Times.}}, \emph{Phys. Rev. Lett.}
  \textbf{52}, (1984) 2313.

\bibitem{Matzner1985}
R.~A. Matzner, C.~Dewitte-morette, and B.~Nelson, \enquote{{Glory scattering by
  black holes}}, \emph{Phys. Rev. D} \textbf{31}, (1985) 1869--1878.

\bibitem{Futterman}
J.~Futterman, F.~Handler and R. A.~
  Matzner, \emph{{Scattering from Black Holes (Cambridge Monographs on
  Mathematical Physics).}} (1988).

\bibitem{Dolan2008a}
S.~R. Dolan, \enquote{{Scattering of long-wavelength gravitational waves}},
  \emph{Phys. Rev. D} \textbf{77}, (2008) 044004.

\bibitem{Dolan2008}
S.~R. Dolan, \enquote{{Scattering and absorption of gravitational plane waves
  by rotating black holes}}, \emph{Class. Quantum Grav.}
  \textbf{25} (2008) 235002.

\bibitem{Crispino2009}
L.~C. Crispino, S.~R. Dolan, and E.~S. Oliveira, \enquote{{Scattering of
  massless scalar waves by Reissner-Nordstr{\"{o}}m black holes}}, \emph{Phys.
  Rev. D} \textbf{79} (2009) 064022.

\bibitem{Crispino2014}
L.~C. Crispino, S.~R. Dolan, A.~Higuchi, and E.~S. {De Oliveira},
  \enquote{{Inferring black hole charge from backscattered electromagnetic
  radiation}}, \emph{Phys. Rev. D} \textbf{90}, (2014) 1--5.

\bibitem{Crispino2015}
L.~C. Crispino, S.~R. Dolan, A.~Higuchi, and E.~S. {De Oliveira},
  \enquote{{Scattering from charged black holes and supergravity}}, \emph{Phys.
  Rev. D} \textbf{92}, (2015) 1--5.

\bibitem{Leite2017}
L.~C.~S. Leite, S.~R. Dolan, and L.~C.~B. Crispino, \enquote{{Absorption of
  electromagnetic and gravitational waves by Kerr black holes: Shadows,
  superradiance and the spin-helicity effect}}, \emph{Phys. Lett. B}
  \textbf{774}, (2017) 1--5.

\bibitem{Alexandre2018}
J.~Alexandre and K.~Clough, \enquote{{Black hole interference patterns in
  flavor oscillations}}, \emph{Phys. Rev. D} \textbf{98} (2018) 043004.

\bibitem{Sporea2018}
C.~A. Sporea, \enquote{{MOG black hole scattering}}, \emph{arXiv:1812.09945} .

\bibitem{Dolan2017}
S.~R. Dolan and T.~Stratton, \enquote{{Rainbow scattering in the gravitational
  field of a compact object}}, \emph{Phys. Rev. D} \textbf{95}, (2017)
  1--20.

\bibitem{Stratton2019}
T.~Stratton and S.~R. Dolan, \enquote{{Rainbow scattering of gravitational
  plane waves by a compact body}}, \emph{arXiv:1903.00025}.

\bibitem{Cotaescu2019}
I.~I. Cot{\u a}escu and C.~A. Sporea, \enquote{{Scattering of Dirac fermions by
  spherical massive bodies}}, \emph{Eur.  Phys. J. C}
  \textbf{79}, (2019) 1--8.

\bibitem{Yoo2013}
C.-M. Yoo, T.~Harada, and N.~Tsukamoto, \enquote{{Wave effect in gravitational
  lensing by the Ellis wormhole}}, \emph{Phys. Rev. D} \textbf{87}, (2013)
  084045.

\bibitem{1988}
K.~Nakajima and H.~Asada, \enquote{{Deflection angle of light in an
    Ellis wormhole geometry}},
  \emph{Phys. Rev. D} \textbf{85}, (2012) 107501.

\bibitem{Sharma2006}
K.~Sharma, \emph{{Optics: principles and applications}} (Academic Press, Tokyo,
  2006).

\bibitem{Ellis1973}
H.~G. Ellis, \enquote{{Ether flow through a drainhole: A particle model in
  general relativity}}, \emph{J. Math. Phys.} \textbf{14}, (1973) 104--118.

\bibitem{Bueno2018}
P.~Bueno, P.~A.~Cano, F.~Goelen, T.~Hertog and B.~Vercnocke, 
\enquote{{Echoes of Kerr-like wormholes}}, \emph{Phys. Rev. D}
\textbf{97}, (2018) 024040.

\end{thebibliography}

\end{document}